\documentclass [aps,pra,amssymb,amsmath,showpacs,twocolumn]{revtex4-1}
\usepackage[latin9]{inputenc}

\pdfoutput=1

\usepackage{times}
\usepackage{amsfonts}
\usepackage{amssymb}
\usepackage{amsmath}

\usepackage{lipsum}
\usepackage{graphicx}

\usepackage[caption=false]{subfig}
\usepackage{bm}
\usepackage{verbatim}
\usepackage[cspex,bbgreekl]{mathbbol}
\allowdisplaybreaks

\usepackage{hyperref}
\hypersetup{
    colorlinks,
    citecolor=blue,
    filecolor=blue,
    linkcolor=blue,
    urlcolor=blue
}

\usepackage{bm} 
\usepackage{color}
\usepackage{stackrel}
\usepackage{accents}

\usepackage{latexsym}

\usepackage{mathtools}

\newcommand{\beg}{\begin{equation}}
\newcommand{\en}{\end{equation}}
\newcommand{\begs}{\begin{subequations}}
\newcommand{\ens}{\end{subequations}}
\newcommand \bea {\begin{eqnarray}}
\newcommand \eea {\end{eqnarray}}
\newcommand{\bem}{\begin{bmatrix}}
\newcommand{\enm}{\end{bmatrix}}
\newcommand{\bpm}{\begin{pmatrix}}
\newcommand{\epm}{\end{pmatrix}}
\newcommand{\bvm}{\begin{vmatrix}}
\newcommand{\evm}{\end{vmatrix}}
\newcommand{\ba}{\begin{array}}
\newcommand{\ea}{\end{array}}

\def\t{\tau}

\def\mean#1{\langle #1 \rangle}

\newcommand{\re}[1]{(\ref{#1})}

\newcommand{\eref}[1]{Eq.~(\ref{#1})}
\newcommand{\fref}[1]{Fig.~\ref{#1}}
\newcommand{\fsref}[1]{Figs.~\ref{#1}}
\newcommand{\Rref}[1]{Ref.~\onlinecite{#1}}

\newcommand{\esref}[1]{Eqs.~(\ref{#1})}

\newcommand{\aref}[1]{Appendix~\ref{#1}}

\def\Z2{\mathbb{Z}_{2}}

\def\R{\mathbb{R}}

\makeatletter
\renewcommand*\env@matrix[1][\arraystretch]{%
  \edef\arraystretch{#1}%
  \hskip -\arraycolsep
  \let\@ifnextchar\new@ifnextchar
  \array{*\c@MaxMatrixCols c}}
\makeatother

\begin{document}

\title{Chaotic Synchronization between Atomic Clocks}

\author{Aniket Patra$^1$, Boris L. Altshuler$^2$ and Emil A. Yuzbashyan$^1$}
\affiliation{$^1$Department of Physics and Astronomy, Rutgers University, Piscataway, NJ 08854, USA \\ 
$^2$Department of Physics, Columbia University, New York, NY 10027, USA}

\begin{abstract} 
 We predict   synchronization of the chaotic dynamics of two atomic ensembles coupled to a heavily damped optical cavity mode. The atoms are dissipated collectively through this mode and pumped incoherently to achieve a macroscopic population of the cavity photons. Even though the dynamics of each ensemble are chaotic, their motions repeat one another. In our system, chaos first emerges via quasiperiodicity and then synchronizes. We identify the signatures of synchronized chaos, chaos, and quasiperiodicity in the experimentally observable power spectra of the light emitted by the cavity.
\end{abstract}

\maketitle


\section{Introduction}

 It is generally  challenging to predict the long term behavior of a chaotic system, e.g., weather, due to  its sensitive dependence on  initial conditions. 
However, there are special chaotic systems where the dynamics of one part are locked or \textit{synchronized} with those of the other part or parts  \cite{Uchida, Chaotic_Synch_1}. As a result,     the asymptotic behavior of certain dynamical variables is fully predictable in spite of the overall chaotic nature. 
   Different mechanisms of obtaining chaotic synchronization have been studied in, for example, electrical circuits \cite{Chaotic_Synch_2, Chaotic_Synch_3}, coupled lasers~\cite{Chaotic_Synch_4, Chaotic_Synch_5, Chaotic_Synch_6},  oscillators in laboratory plasma~\cite{Chaotic_Synch_7}, population dynamics \cite{Chaotic_Synch_8} and earthquake models \cite{Chaotic_Synch_9}. 
   In this paper, we report chaotic synchronization in a novel physical system, namely, between two \textit{mutually coupled} ensembles of atoms   in a driven-dissipative experimental setup. This is unlike most other examples of chaotic synchronization, where the coupling between the two parts is  unidirectional.   In the chaotic synchronized phase, our system has potential applications in secure communication \cite{Uchida, Chaotic_Synch_1, Chaotic_Synch_2, Chaotic_Synch_3, Chaotic_Synch_6}. 

\begin{figure}[tbp!]
\begin{center}
\includegraphics[scale=0.45]{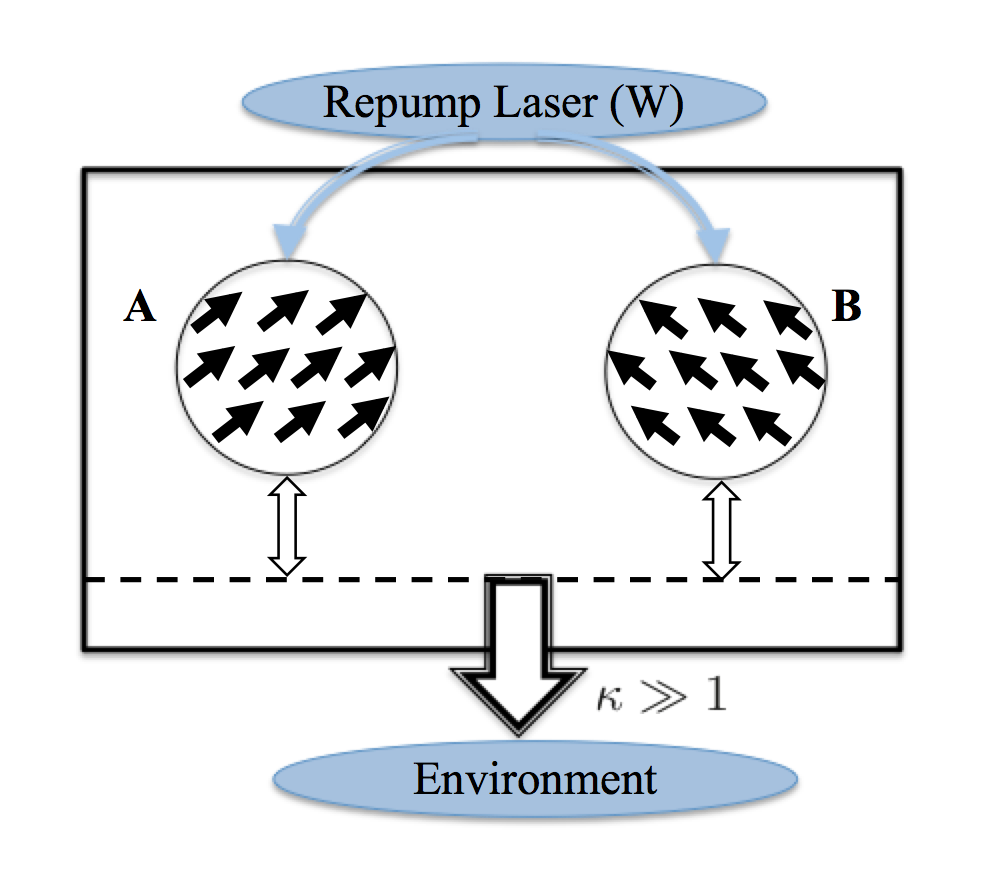}
\caption{Cartoon depicting the driven-dissipative experimental setup with two atomic ensembles inside a bad cavity. In ensembles `A' and `B', the solid arrows denote individual atoms. The thick double-headed arrows correspond to the Rabi coupling between the ensembles and the cavity mode (dashed line). The rate of loss of photons from the cavity is $\kappa$.}\label{Setup}
\end{center}
\end{figure}    
  
\begin{figure}[tbp!]
\begin{center}
\includegraphics[scale=0.46]{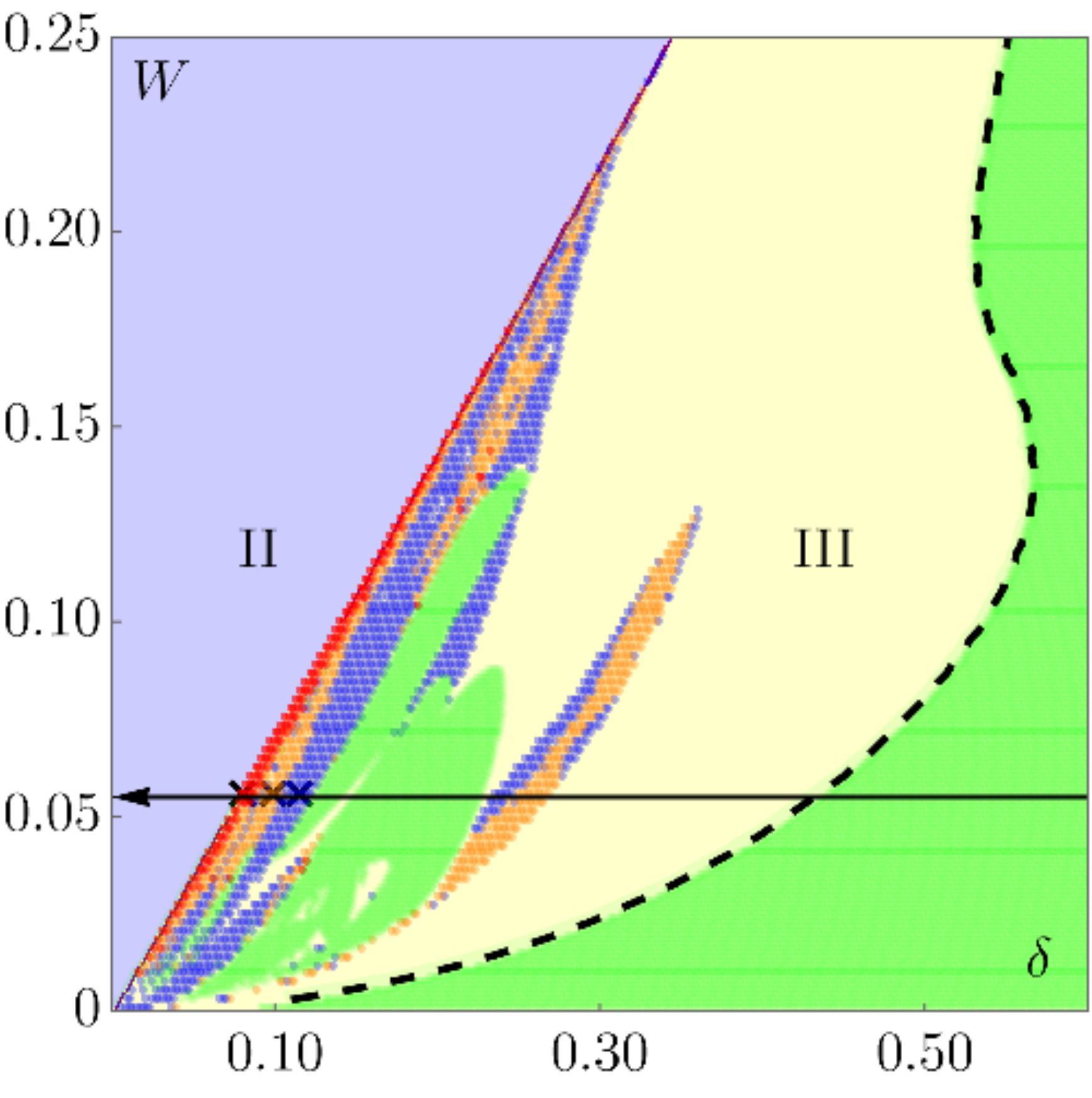}
\begin{picture}(0,0)
\put(-85,140){\includegraphics[height=3cm]{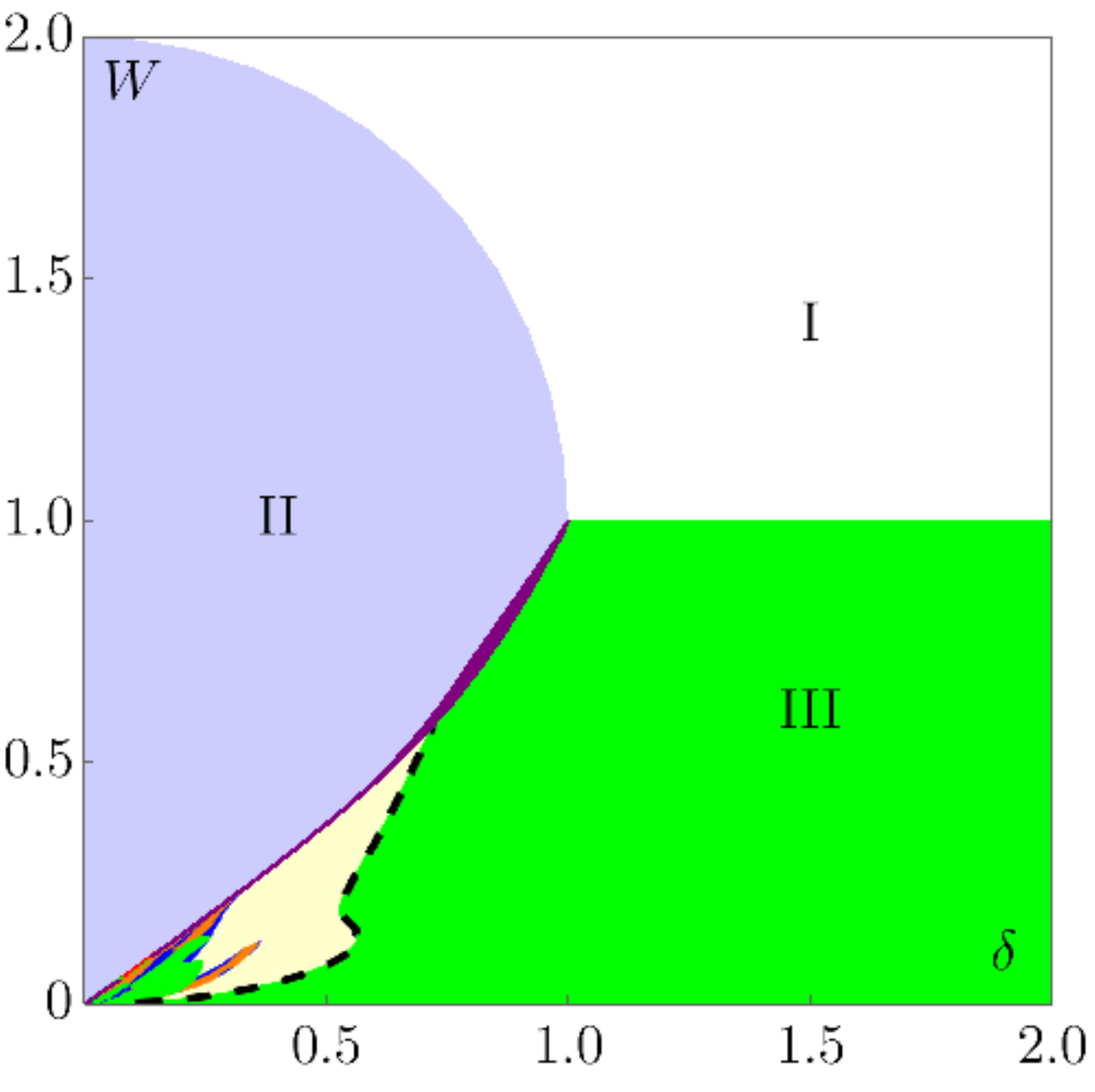}}
\end{picture}
\caption{(color online) Nonequilibrium phase diagram for two atomic ensembles in a bad optical cavity.  $W$ is the repump rate and and $\delta$ is the detuning between the atomic level spacings   of the two ensembles  in  the units of the collective decay rate $N\Gamma_c$.  The inset  shows the full phase diagram with Phases I (normal, non-superradiant    phase), II (monochromatic superradiance), and III (amplitude-modulated superradiance). The main picture is a blowup of the region near the origin. Green points correspond to $\Z2$-symmetric (with respect to the interchange of the two ensembles) collective oscillations (limit cycle). The $\Z2$ symmetry breaks spontaneously across the black dashed line. In the yellow region to the left of this line, the  attractor is a symmetry-broken limit cycle. Dark blue, orange, and red points  indicate quasiperiodicity, chaos, and synchronized chaos, respectively. }
\label{Near_Origin}
\end{center}
\end{figure} 

\begin{figure*}[tbp!]
\centering
\subfloat[\large (a)]{\label{QP_ZA}\includegraphics[scale = 0.30]{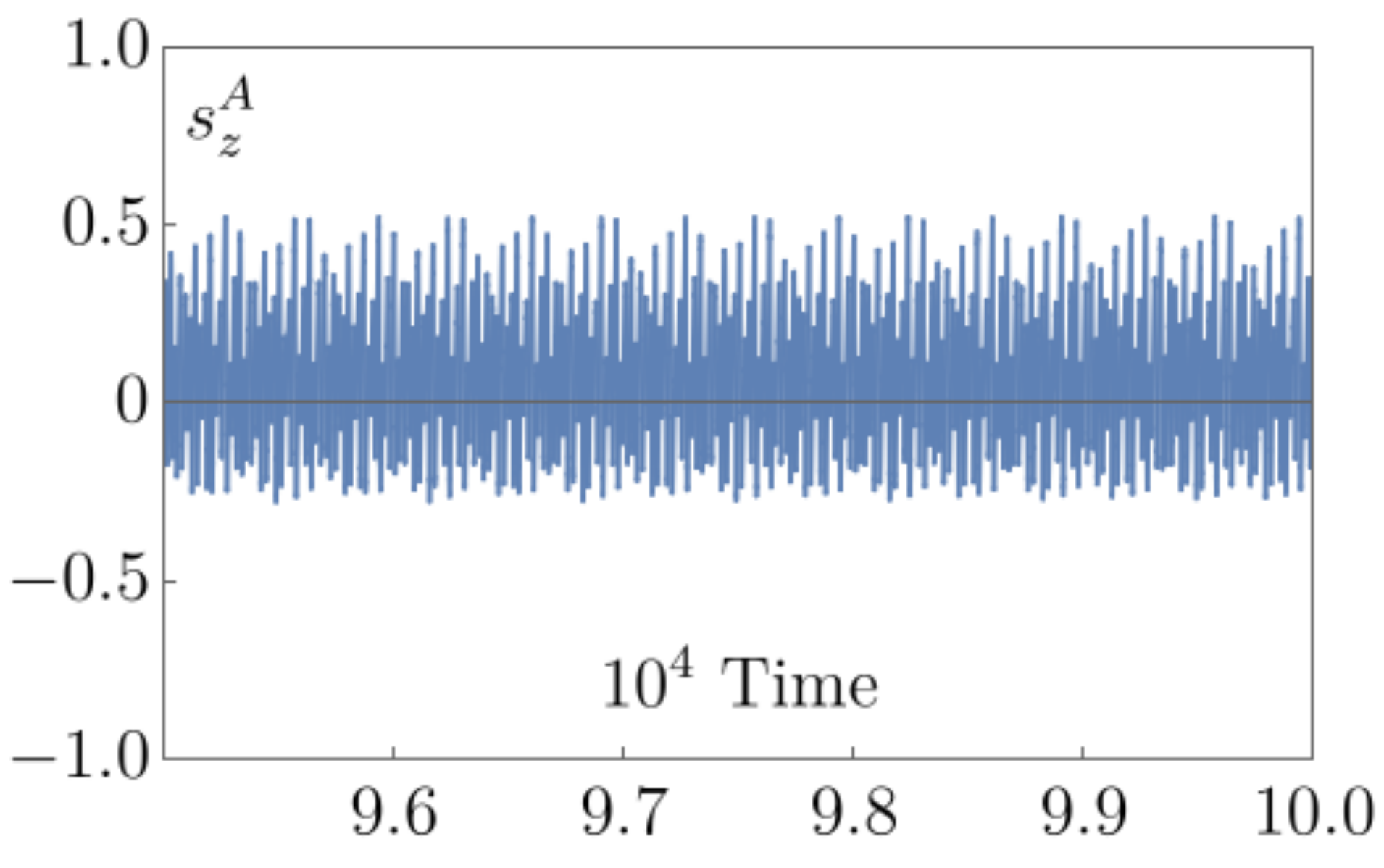}}\quad
\subfloat[\large (b)]{\label{QP_SZA_vs_SZB}\includegraphics[scale = 0.185]{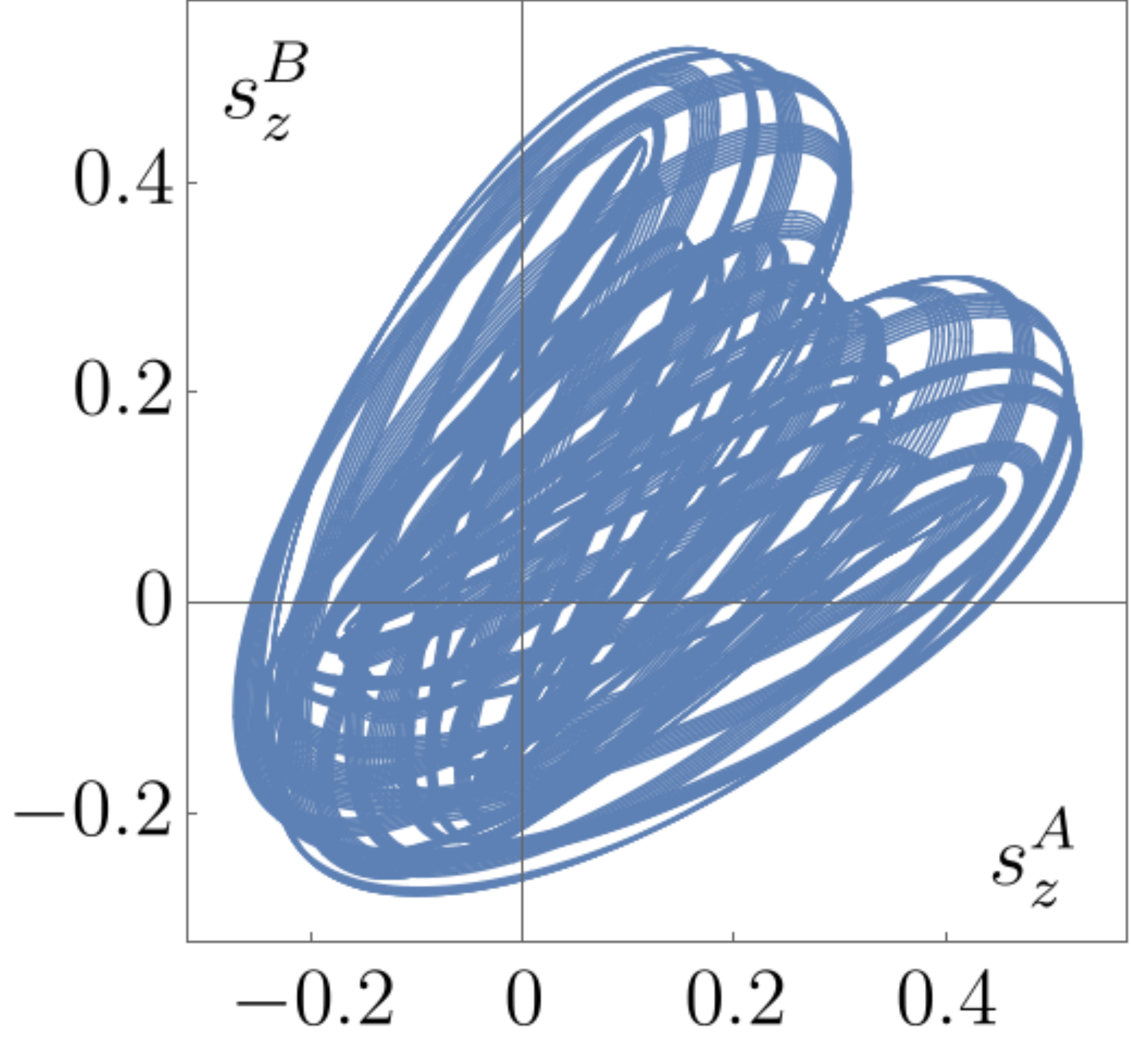}}\qquad
\subfloat[\large (c)]{\label{Spectrum_QP}\includegraphics[scale = 0.30]{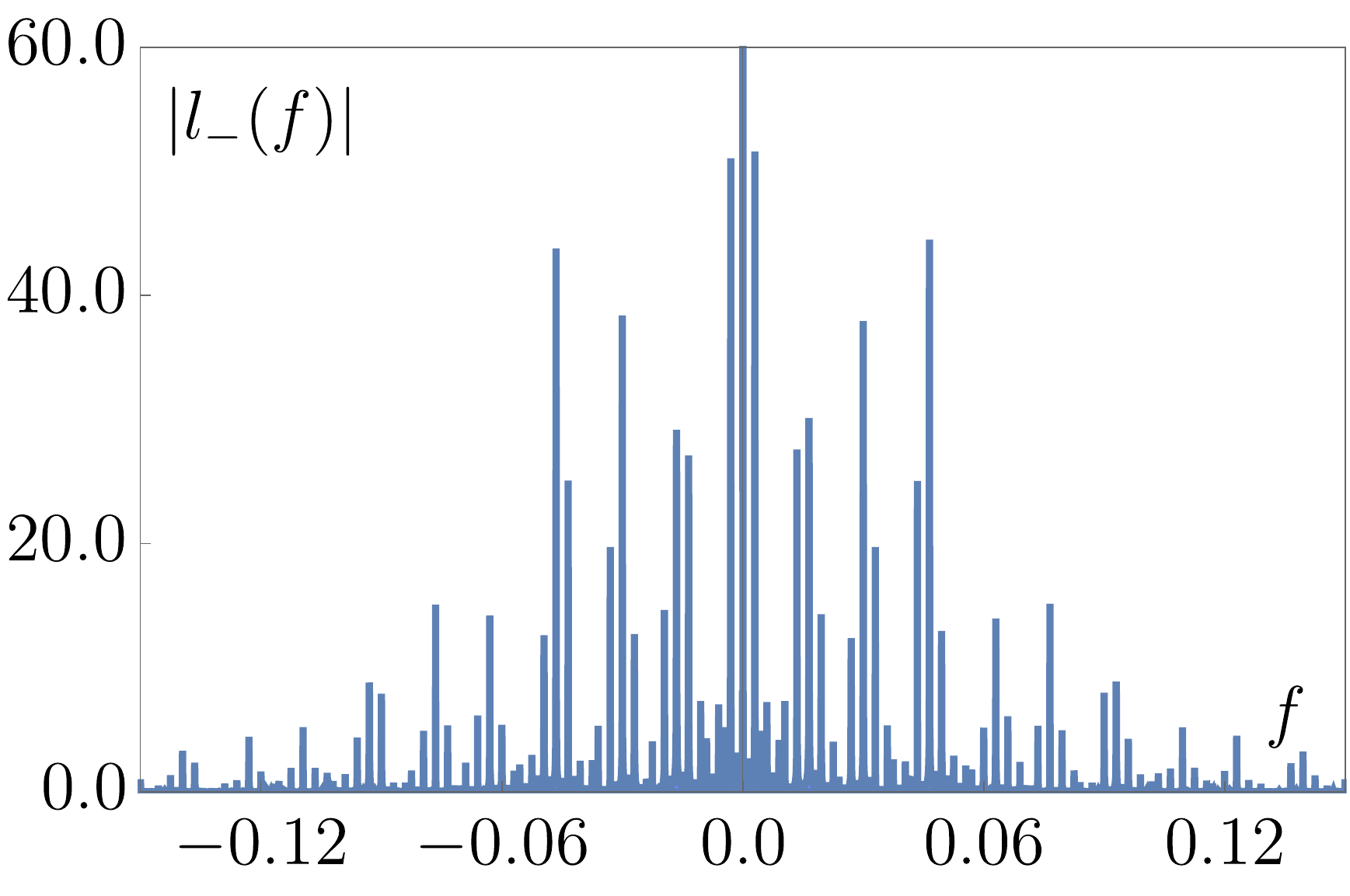}}\\
\subfloat[\large (d)]{\label{C_ZA}\includegraphics[scale = 0.30]{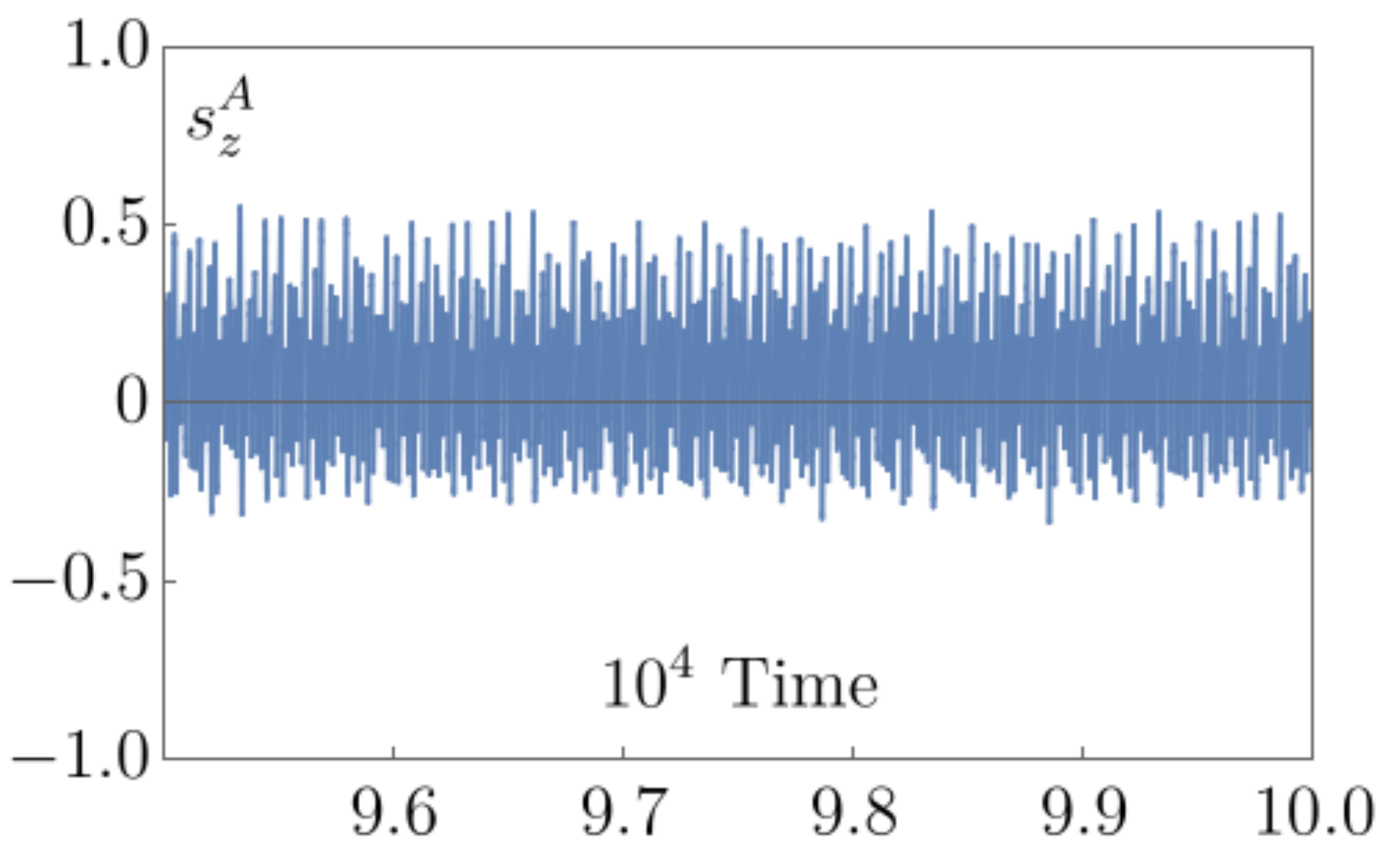}}\quad
\subfloat[\large (e)]{\label{C_SZA_vs_SZB}\includegraphics[scale = 0.185]{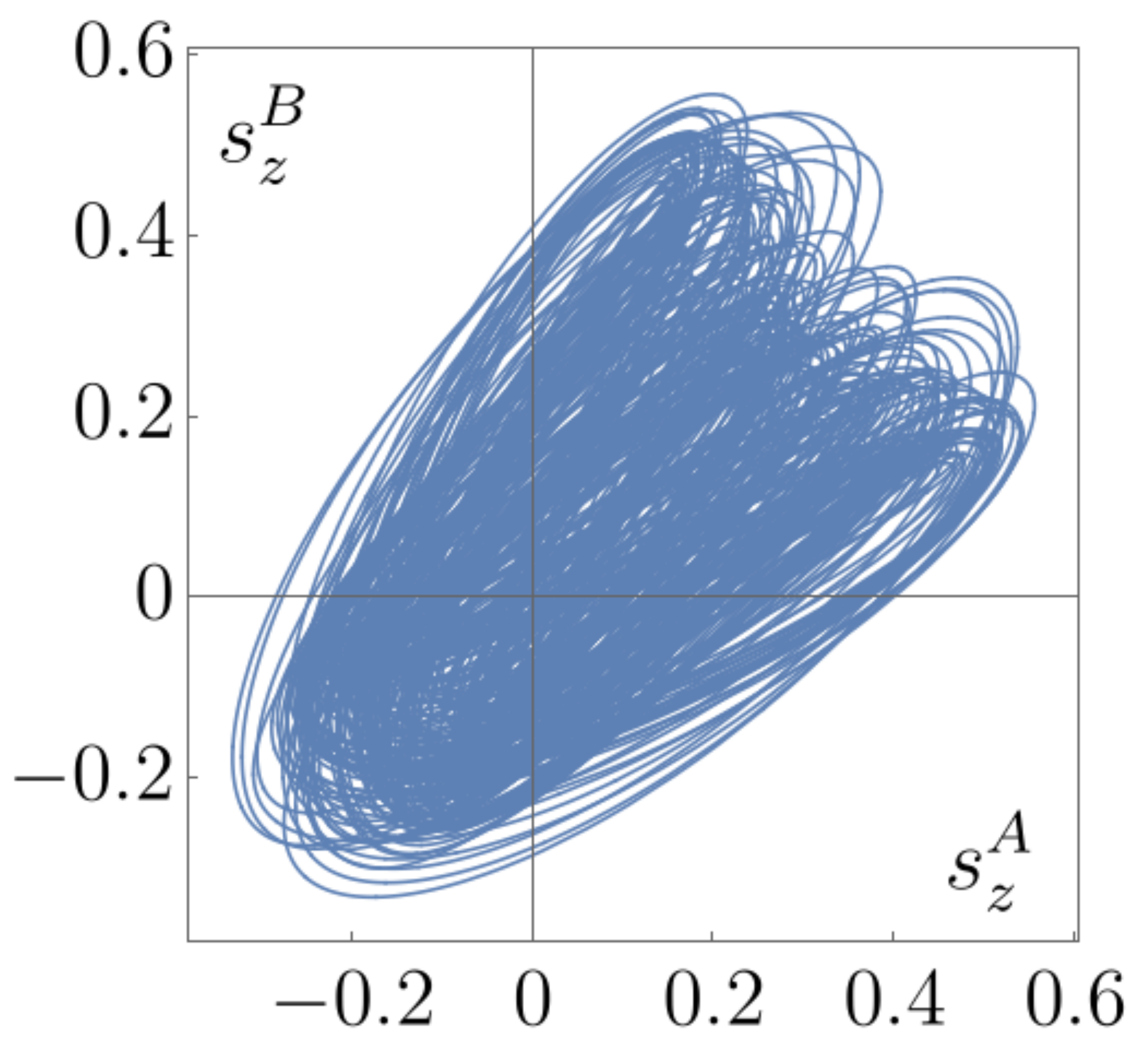}}\qquad
\subfloat[\large (f)]{\label{Spectrum_C}\includegraphics[scale = 0.30]{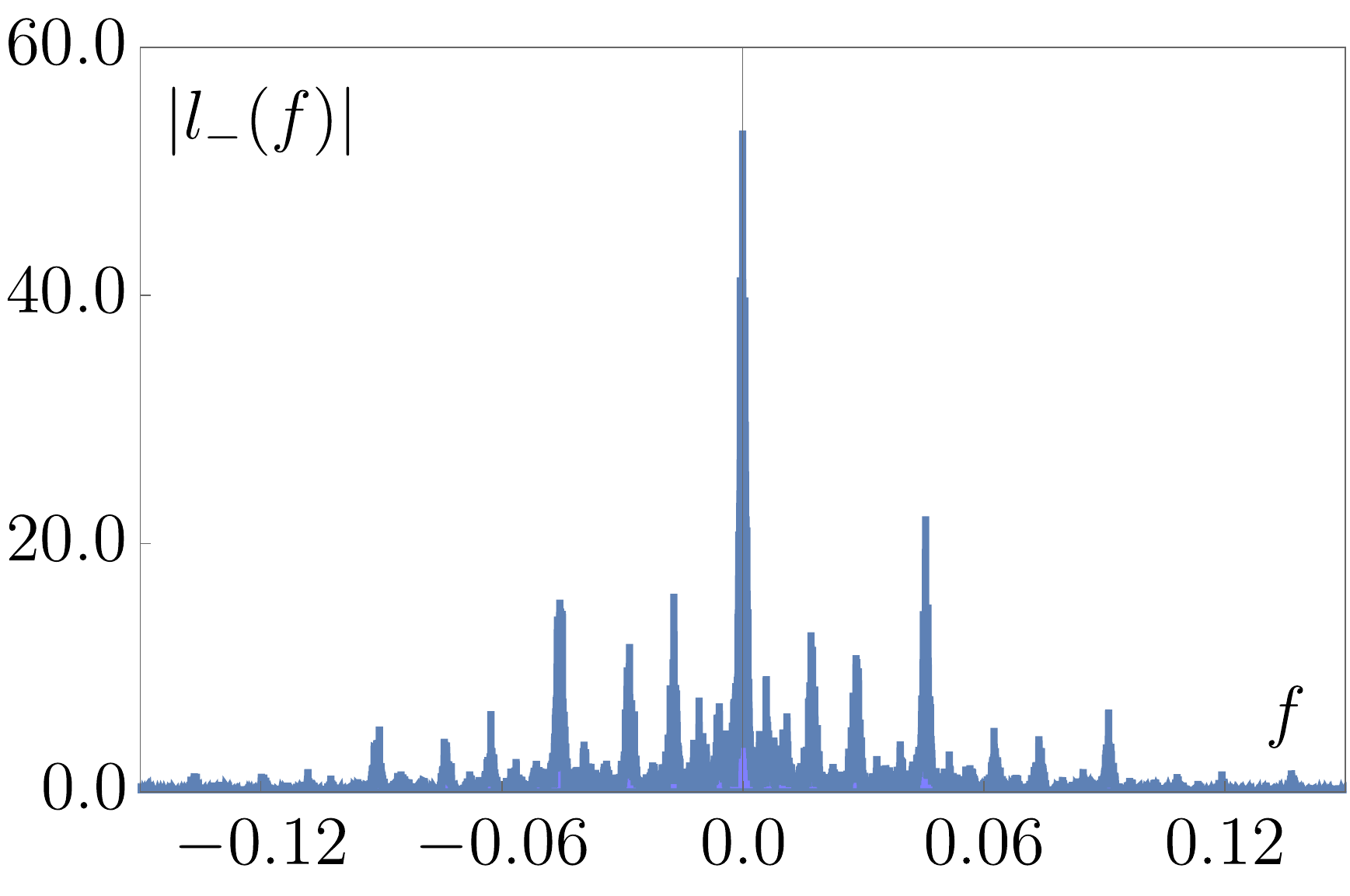}}\llap{\raisebox{1.95cm}{\includegraphics[height=1.5cm]{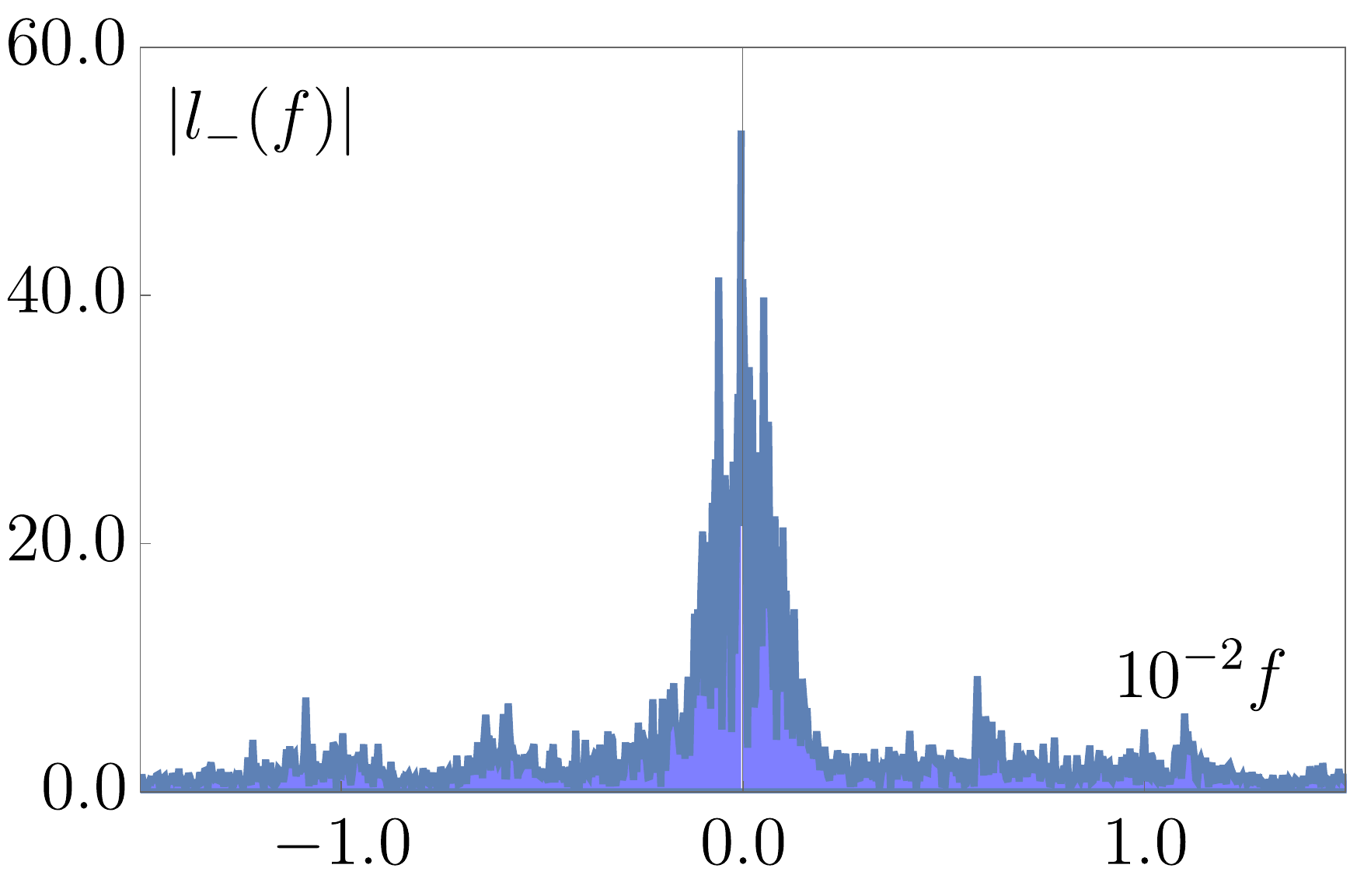}}}\\
\subfloat[\large (g)]{\label{SC_ZA}\includegraphics[scale = 0.30]{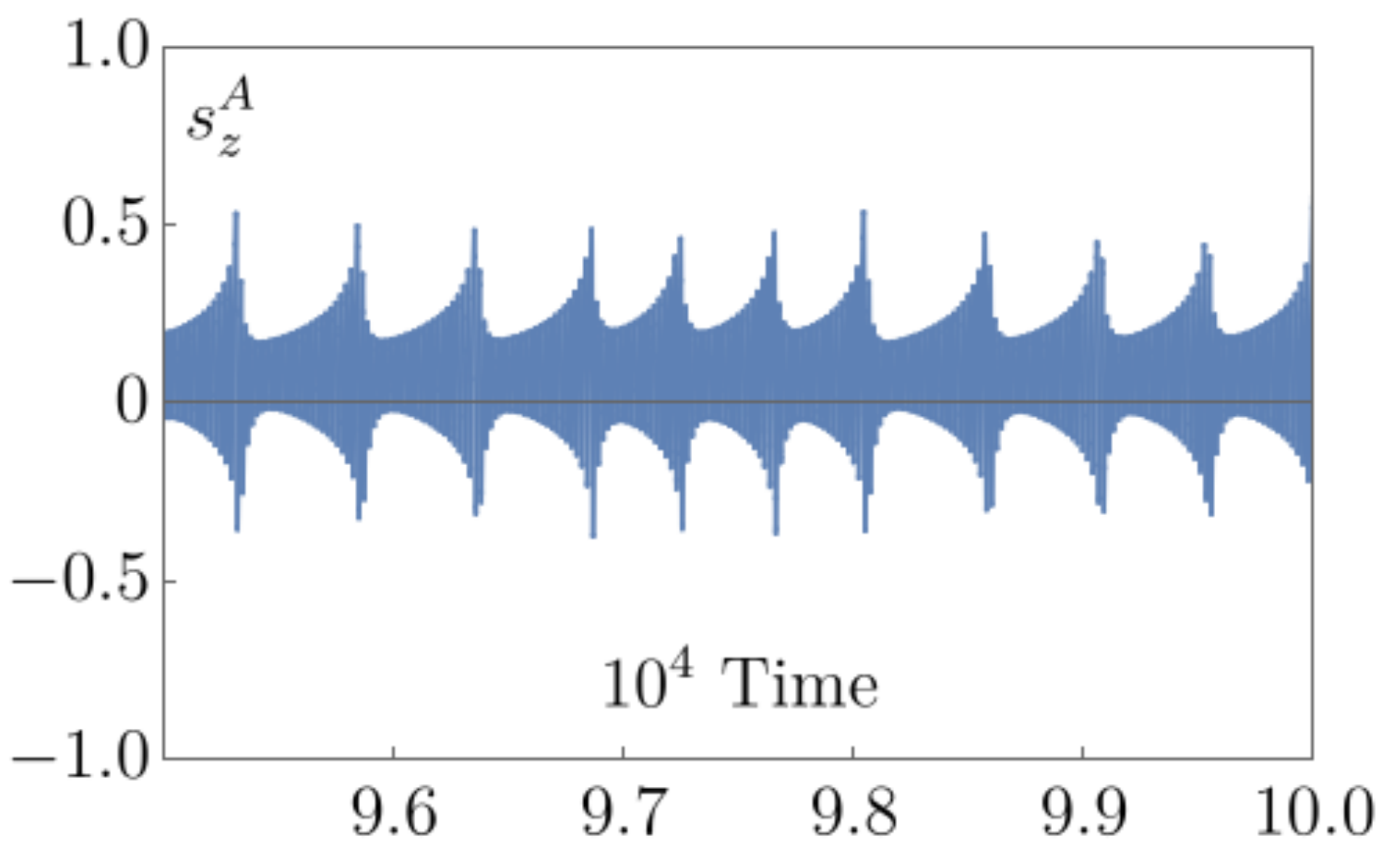}}\quad
\subfloat[\large (h)]{\label{SC_SZA_vs_SZB}\includegraphics[scale = 0.185]{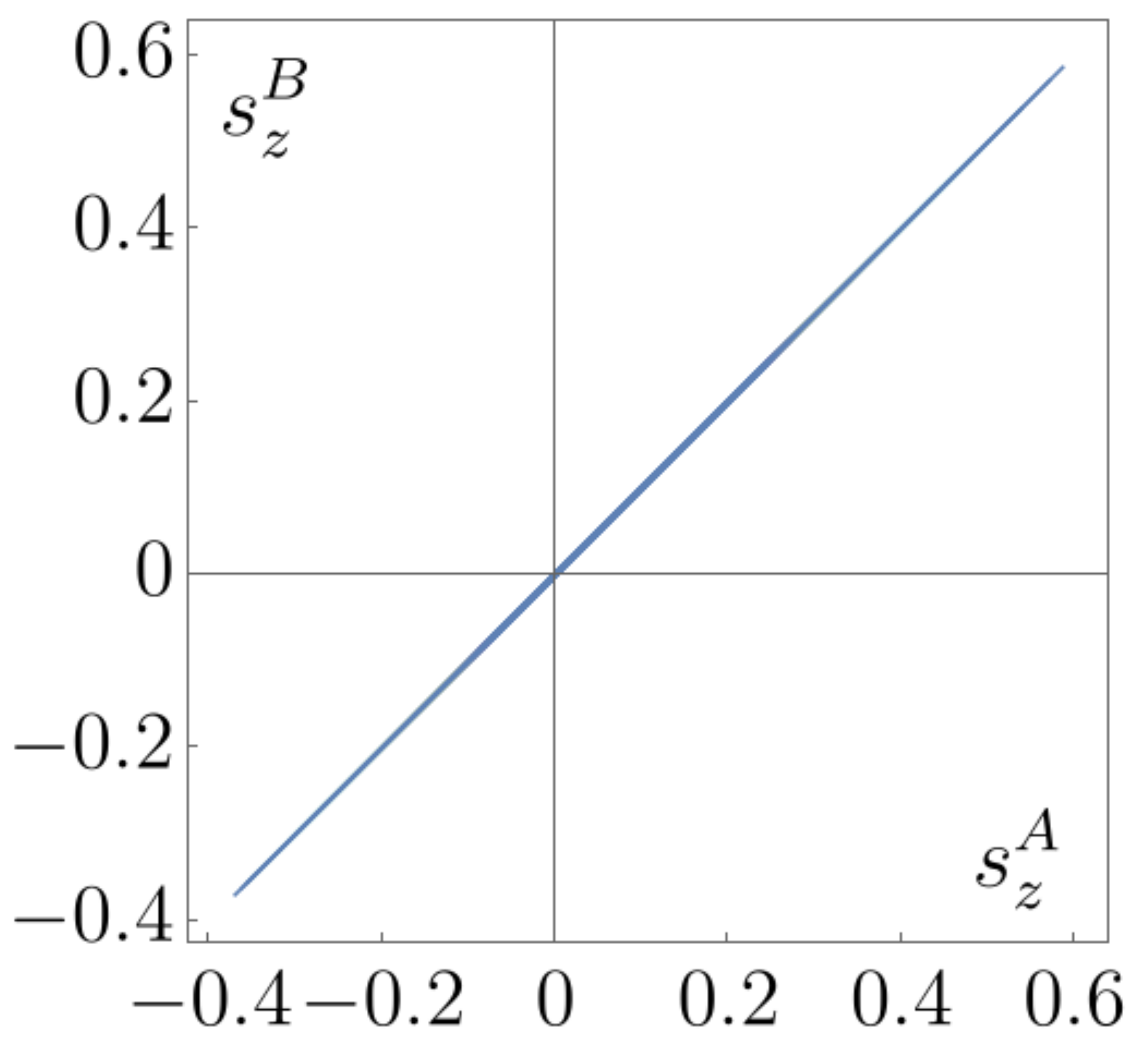}}\qquad
\subfloat[\large (i)]{\label{Spectrum_SC}\includegraphics[scale = 0.30]{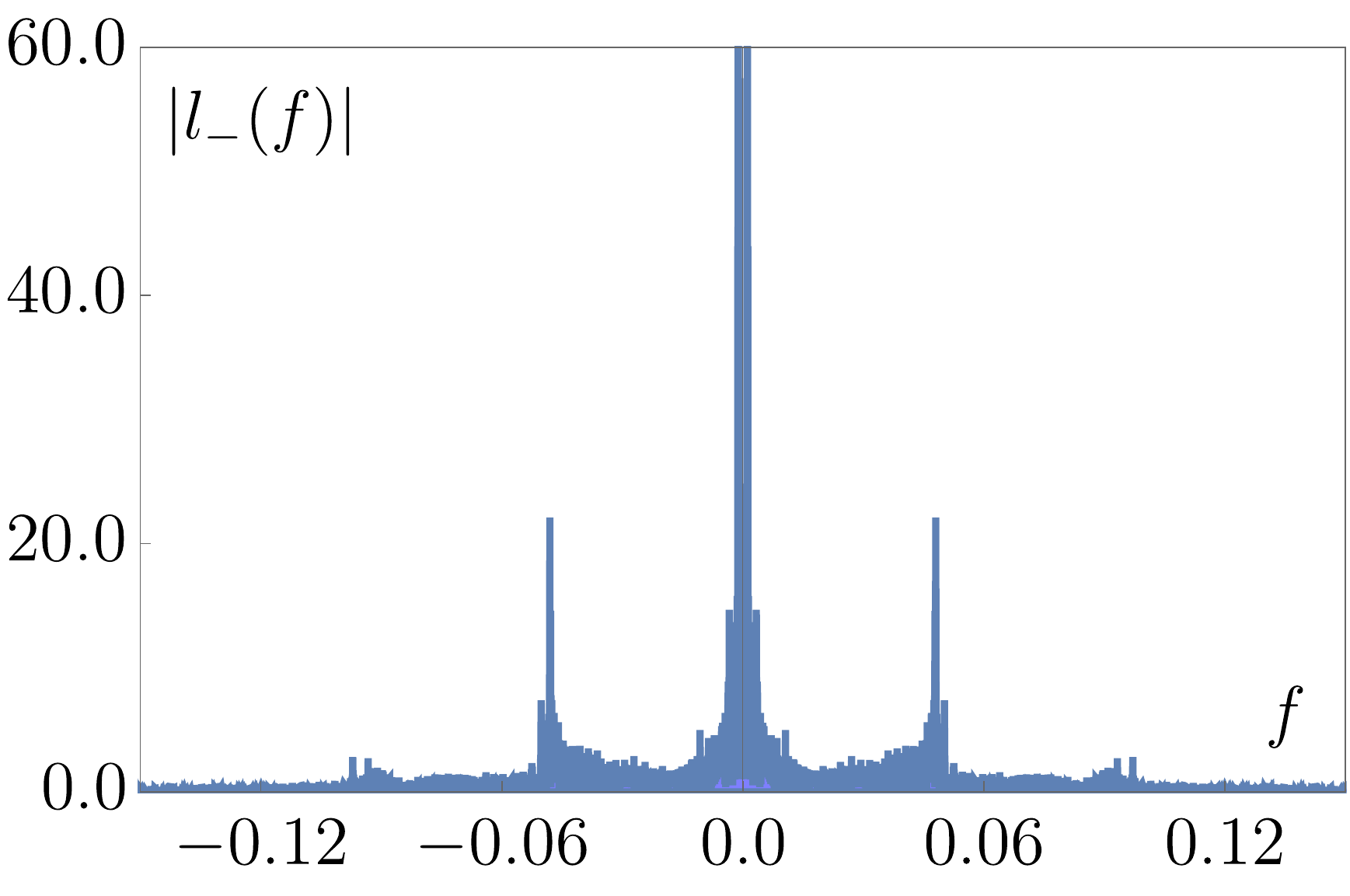}}\llap{\raisebox{1.95cm}{\includegraphics[height=1.5cm]{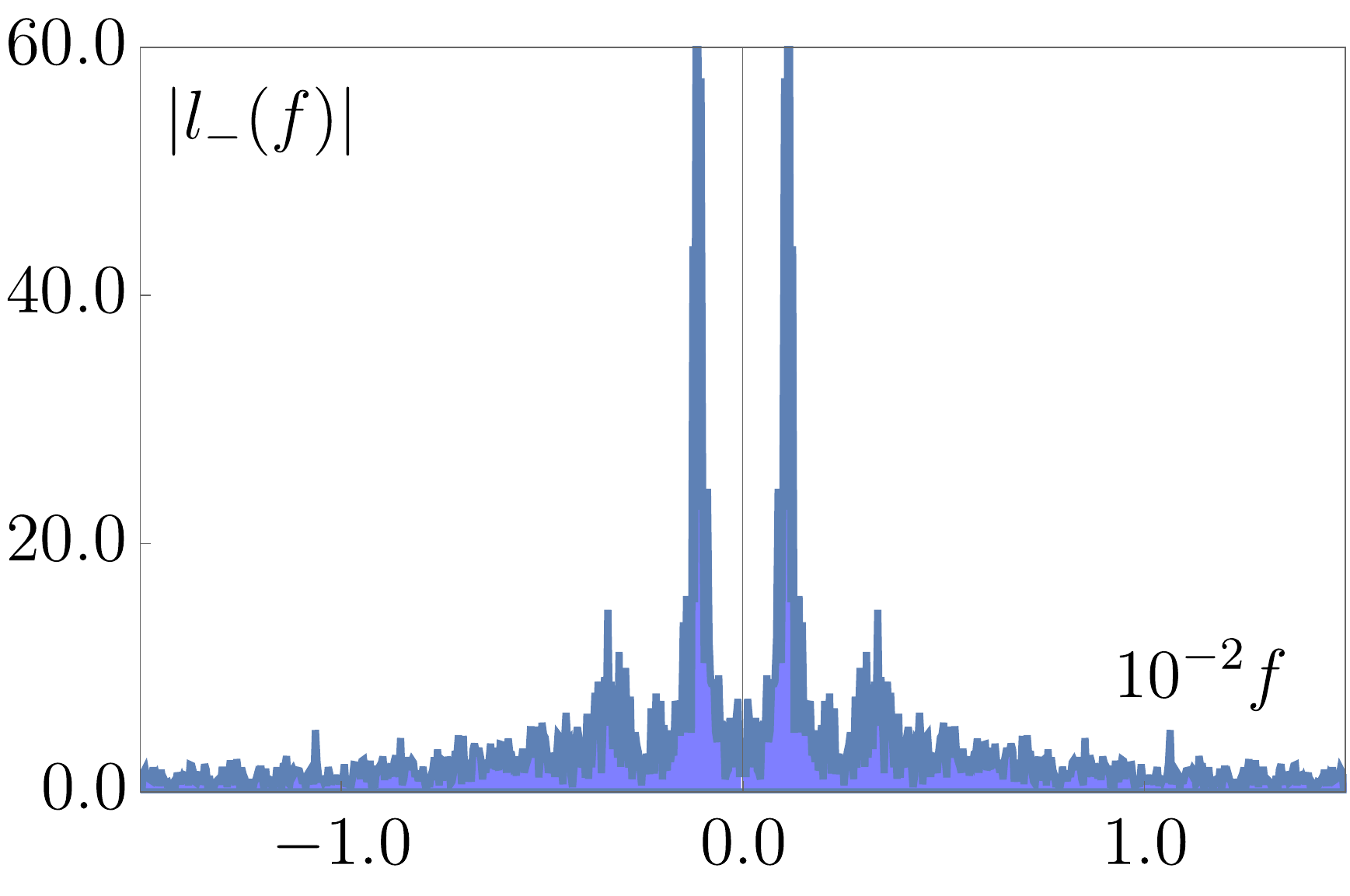}}}
\caption{Three   types of dynamics of two coupled atomic clocks, i.e., of two atomic ensembles   coupled to a strongly damped cavity mode. The classical spins $\bm{s}^{\tau}$  representing the two clocks ($\tau = A$ and $B$) obey   mean-field equations of motion \re{Mean-Field_1}.  The three rows of plots from top to bottom  represent  quasiperiodic ($\delta = 0.115, W = 0.055$),    chaotic ($\delta = 0.1, W = 0.055$), and  synchronized chaotic ($\delta = 0.080, W = 0.055$) attractors, respectively.  We marked these three $(\delta, W)$ pairs   with crosses   on the solid black horizontal arrow in \fref{Near_Origin}.   The three columns of plots from left to right show  $s_{z}^{A}$ vs. time [\fsref{QP_ZA}, \ref{C_ZA}, and \ref{SC_ZA}], $s_{z}^{B}$ vs. $s_{z}^{A}$ [\fsref{QP_SZA_vs_SZB}, \ref{C_SZA_vs_SZB}, and \ref{SC_SZA_vs_SZB}], and the  power spectra of the radiated light [\fsref{Spectrum_QP}, \ref{Spectrum_C}, and \ref{Spectrum_SC}], respectively. In the quasiperiodic spectrum \textbf{(c)}, the main peaks are at integer multiples of $f_{1} \approx 1.6 \times 10^{-2}$, whereas the distance between  auxiliary peaks is $f_{2} \approx 3.0 \times 10^{-3}$.  In the insets to plots \textbf{(f)} and \textbf{(i)} we magnified the region near $f = 0$ to show the presence  or  absence of the peak at the origin. } \label{Spectrum_C_Full}
\end{figure*}

We consider two spatially separated ensembles of, e.g., $\prescript{87}{}{\textrm{Rb}}^{}_{}$ atoms inside a  bad (leaky)  optical cavity, see \fref{Setup}. The atoms are collectively dissipated through a Rabi coupling to a   heavily damped  cavity mode and pumped with a transverse laser to achieve a macroscopic population of the  cavity photons  \cite{ Holland_Two_Ensemble_Expt}. A single atomic ensemble  coupled to a bad cavity has been proposed as a source of ultracoherent  radiation for an  atomic clock~\cite{Holland_atomic_clock}. The two ensembles in our setup, therefore, represent two  interacting atomic clocks~\cite{Holland_Two_Ensemble_Theory}.
Previous work  obtained  main nonequilibrium phases  of this system \cite{Holland_Two_Ensemble_Theory,Patra_1}, see the inset in \fref{Near_Origin}. 

In this paper, we study a finite region of the phase diagram -- the orange (nonsynchronized chaos) and  red (synchronized chaos) points in region III of \fref{Near_Origin} -- where the light radiated  by the cavity behaves chaotically. Here chaos appears via quasiperiodicity \cite{Hilborn, QP_Chaos_1, QP_Chaos_2}. Initially,  chaotic trajectories fill up extended regions in the configuration space, see \fref{C_SZA_vs_SZB}. We discover a subregion inside the chaotic phase where  dynamics are confined to a flat hypersurface (\fref{SC_SZA_vs_SZB}), called the ``synchronization manifold". Essentially, the time dependence of one ensemble follows that of the other. We also study  signatures of  these novel behaviors in the   power spectra  of the radiated light. Unlike the quasiperiodic power spectrum (\fref{Spectrum_QP}), which consists of discrete peaks, the chaotic  one (\fref{Spectrum_C}) is a continuum.  The chaotic synchronized spectrum (\fref{Spectrum_SC}) additionally has a reflection symmetry about zero   and no peak at zero frequency, see the insets in \fsref{Spectrum_C} and \ref{Spectrum_SC}.

\section{Nonchaotic Phases}

 We model our system (cf. \fref{Setup}) with the following master equation for the density matrix $\rho$:
\begs
\bea
\dot{\rho} &=& -\imath \big[\hat{H}, \rho\big] + \kappa\mathcal{L}[a]\rho + W\sum_{\tau = A,B}\sum_{j = 1}^{N}\mathcal{L}[\hat{\sigma}^{\tau}_{j+}]\rho, \\ \label{Master} 
\hat{H} &=& \omega_{0}\hat{a}^{\dagger}\hat{a} + \!\!\! \sum_{\tau = A,B}\big[\omega_{\tau}\hat{S}_{\tau}^{z} + \frac{\Omega}{2}\big(\hat{a}^{\dagger}\hat{S}_{\tau}^{-}+\hat{a}\hat{S}_{\tau}^{+}\big)\big]. \label{H}  
\eea 
\label{Full_Master}%
\ens        
The Hamiltonian  $\hat{H}$ describes two atomic ensembles A and B Rabi coupled (with frequency $\Omega$) to the cavity mode $\omega_{0}$, where $\hat{a}^{\dagger} (\hat{a})$ create (annihilate) cavity photons. Each   ensemble contains a large number of atoms, e.g., $N \approx 10^{6}$ of $\prescript{87}{}{\textrm{Rb}}^{}_{}$ atoms  \cite{Holland_Two_Ensemble_Theory, Holland_Two_Ensemble_Expt}. We focus on the lasing transition between two atomic levels. Consequently, we describe individual  atoms with Pauli matrices and   the atomic ensembles with collective spin operators $\hat{S}^{A,B}_{z} = \frac{1}{2}\sum_{j = 1}^{N}\hat{\sigma}^{A,B}_{jz}$ and $\hat{S}^{A,B}_{\pm} = \sum_{j = 1}^{N}\hat{\sigma}^{(A,B)}_{j\pm}$. Experimentally, the level-spacings $\omega_{\tau}$ are controlled  with two   distinct Raman dressing lasers \cite{Holland_Two_Ensemble_Expt}. We model the energy nonconserving processes (decay of the bad cavity mode with a rate $\kappa (\gg 1)$, and incoherent pumping by external lasers at an effective repump rate $W$) by  Lindblad superoperators, 
\beg 
\mathcal{L}[\hat{O}]\rho \equiv \frac{1}{2}\big(2\hat{O}\rho \hat{O}^{\dagger} - \hat{O}^{\dagger}\hat{O}\rho -\rho\hat{O}^{\dagger}\hat{O}\big).
\label{Lindblad_def}
\en%
Using the adiabatic approximation \cite{Bad_cavity}, which is exact in the limit $\kappa \rightarrow \infty$, we eliminate the cavity mode replacing $\hat{a}\to \frac{\imath \Omega}{\kappa}\sum_{\tau} \hat{S}_{\tau}^{-}$. Finally, in the rotating frame, where  frequencies are shifted by the mean level-spacing  (it is equal to the clock transition frequency, which is $\approx 6.8$  GHz for $\prescript{87}{}{\textrm{Rb}}^{}_{}$), we derive  semiclassical evolution equations using the mean-field approximation, $\mean{\hat{O_{1}}\hat{O_{2}}} \approx \mean{\hat{O_{1}}}\mean{\hat{O_{2}}}$,
\begs
\bea
\dot{s}^{\t}_{\pm} &=& \biggl(\pm\imath \omega_{\t}- \frac{W}{2}\biggr)s^{\tau}_{\pm} + \frac{1}{2}s^{\t}_{z}l_{\pm}, \\ \label{REOMPM}
\dot{s}^{\t}_{z} &=&  W\big(1 - s^{\t}_{z}\big) - \frac{1}{4}s^{\t}_{+}l_{-} - \frac{1}{4}s^{\tau}_{-}l_{+}, \label{REOMz} 
\eea 
\label{Mean-Field_1}%
\ens%
where $\tau = A, B$, $s^{\t}_{\pm} = \frac{2}{N}\big(\mean{\hat{S}_{x}^{\tau}} \pm i\mean{\hat{S}_{y}^{\tau}}\big)$, $s^{\t}_{z} = \frac{2}{N}\mean{\hat{S}_{z}^{\tau}}$, $\bm{l} = \frac{2}{N}\big(\mean{\hat{\bm{S}}^{A}}+\mean{\hat{\bm{S}}^{B} }\big)$ is the total classical spin,  $\omega_{A} = \delta/2$ and $\omega_{B} = -\delta/2$.   In  \eref{Mean-Field_1} and from now on, we express the detuning $\delta$ and the repump rate $W$ in the units of the collective decay rate $N\Gamma_c\equiv\frac{N\Omega^{2}}{\kappa}$ ($\approx1.4$ kHz for a typical experimental setup \cite{Holland_Two_Ensemble_Expt}) and replace $ (N\Gamma_c) t\to t$. 

The mean-field equations of motion possess   two symmetries: $\bf (1)$    Axial symmetry $\bm{s}^{\tau}\to\mathbb{R}(\phi)\cdot\bm{s}^{\tau}$, where   $\mathbb{R}(\phi)$  is a rotation by an angle $\phi$  around the $z$-axis.  Indeed, the replacement $s_{\pm}^{\tau}\to s_{\pm}^{\tau}e^{\pm \imath \phi}$  leaves   \eref{Mean-Field_1} unchanged. $\bf (2)$    $\Z2$ symmetry $\bm{s}^{\tau}\to \mathbb{\Sigma}\circ \R(\phi_{0}) \cdot \bm{s}^{\tau}$ which involves a rotation around the $z$-axis by a fixed angle $\phi_{0}$ followed by an interchange, 
\beg  
\mathbb{\Sigma}:\,\big( s_{\pm}^{A}, s_{z}^{A}, s_{\pm}^{B}, s_{z}^{B} \big) \longrightarrow \big( s_{\mp}^{B}, s_{z}^{B}, s_{\mp}^{A}, s_{z}^{A} \big), 
\label{Z2}
\en%
of ensembles A and B while flipping  the sign of $s_{y}$.
 The $\Z2$ symmetric solutions  obey $\bm{s}^{\tau} = \mathbb{\Sigma}\circ \R(\phi_0) \cdot \bm{s}^{\tau}$, where the value of  $\phi_{0}$ depends on the initial condition. This constraint  defines a  4D $\Z2$-symmetric submanifold.  
 All attractors in \fref{Near_Origin}, except the normal phase, spontaneously break the axial symmetry. This implies that for each $(\delta, W)$
 point there is a family of attractors related by a rotation $\mathbb{R}(\phi)$ around the $z$-axis, where $\phi$ depends on the initial condition.
 
 The limit cycle (periodically modulated superradiance)   in the green region of  \fref{Near_Origin} possesses  $\Z2$ symmetry, which breaks spontaneously across the black dashed line. In the absence of any symmetry, the interaction between the two spins introduces additional frequencies. The ratio of two such frequencies being irrational causes quasiperiodicity, which eventually gives way to chaos. At its inception, the chaotic attractor is completely asymmetric. As we decrease $\delta$ while keeping $W$ fixed, one spin gets locked to the other. We  interpret this synchronized chaotic phase as  spontaneous restoration of the $\Z2$ symmetry.  In this phase,   the   conditional Lyapunov exponent  becomes negative, while the maximum Lyapunov exponent  remains positive \cite{Uchida, Chaotic_Synch_1}  as shown in \fref{Cond_Lya}. As we cross over  to region II, chaos disappears altogether. One is left with monochromatic superradiance, which  is a fixed point of \eref{Mean-Field_1} \cite{Patra_1}. Separately, we also   note that dynamics of a single atomic ensemble coupled to a bad cavity show no chaos or quasiperiodicity. In fact, this case corresponds to $\delta=0$ in \eref{Mean-Field_1}, i.e., to the vertical axis of the phase diagram in \fref{Near_Origin}, where only Phases I and II are present \cite{Patra_1}.
 
 \section{Synchronization of Chaos}
 
\begin{figure}[tbp!]
\begin{center}
\includegraphics[scale=0.5]{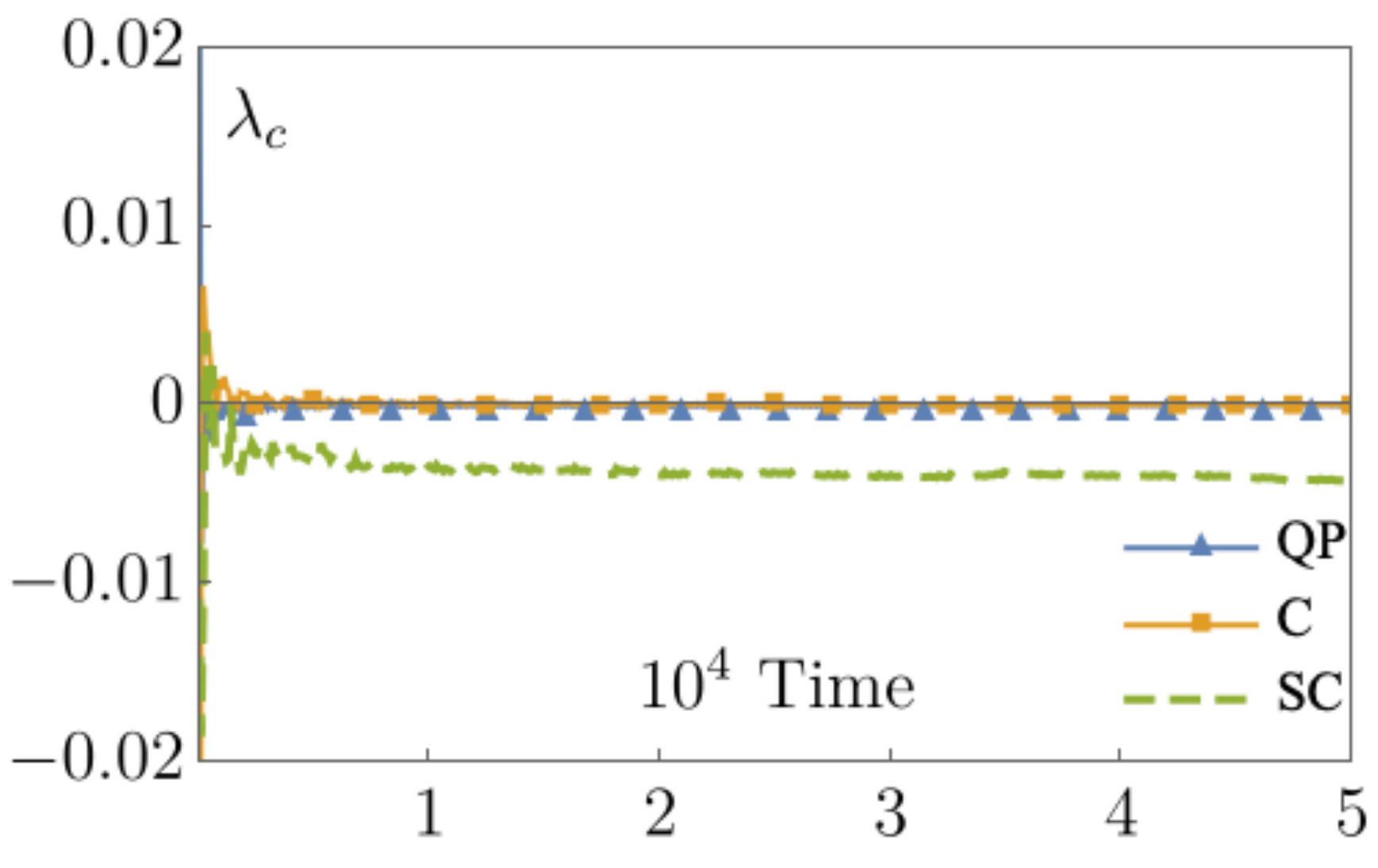}
\begin{picture}(0,0)
\put(12,103){\includegraphics[height=2.5cm]{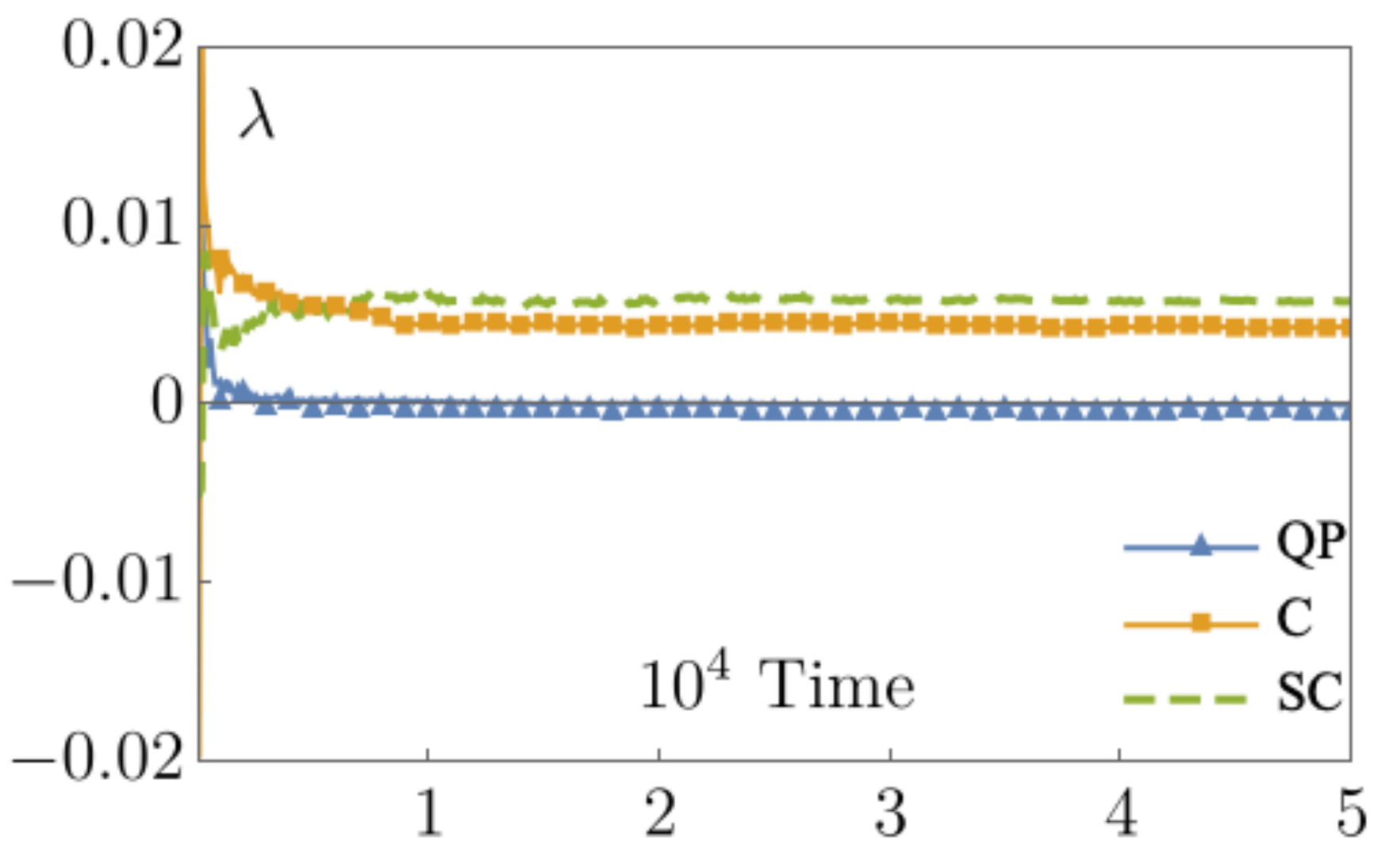}}
\end{picture}
\caption{(color online) Conditional Lyapunov exponents $\lambda_c(t)$   quickly saturate to zero for quasiperiodic (blue triangles) and chaotic (yellow squares) attractors. The values of the detuning $\delta$ and the repump rate $W$ are the same as in \fref{Spectrum_C_Full}. For synchronized chaos (green dashed line)  $\lambda_{c}$ is negative. The inset shows the maximum Lyapunov exponent $\lambda(t)$.  As expected, for chaos and synchronized chaos $\lambda$ is positive, whereas for quasiperiodicity  it saturates to zero.}\label{Cond_Lya}
\end{center}
\end{figure}

 In the rest of this paper, we analyze the evolution from quasiperiodicity to synchronized chaos with the help of Poincar\'{e} sections, Lyapunov exponents, and power spectra.
 We   define the maximum Lyapunov exponent $\lambda(t)$ as usual,
  \beg
  \lambda(t) = \lim_{ d(0)\to 0} \frac{1}{t}\ln\left[\frac{d(t)}{d(0)}\right],
  \label{lyapunov}
  \en
  where $d(t)$ is the distance in the 6D real space of six components of vectors $\bm{s}^{A}$ and $\bm{s}^{B}$.  The conditional Lyapunov exponent is the maximum Lyapunov exponent for  directions transverse to the synchronization manifold, see \eref{Trans_Coord}. For both chaotic and synchronized chaotic attractors $\lambda(t)$ converges to a positive value  ($\approx10^{-2} \pm 10^{-5}$), whereas for the quasiperiodic attractor it  vanishes  ($\pm 10^{-5}$), see the inset in \fref{Cond_Lya}.

\begin{figure*}[tbp!]
\begin{center}
\includegraphics[scale=0.5]{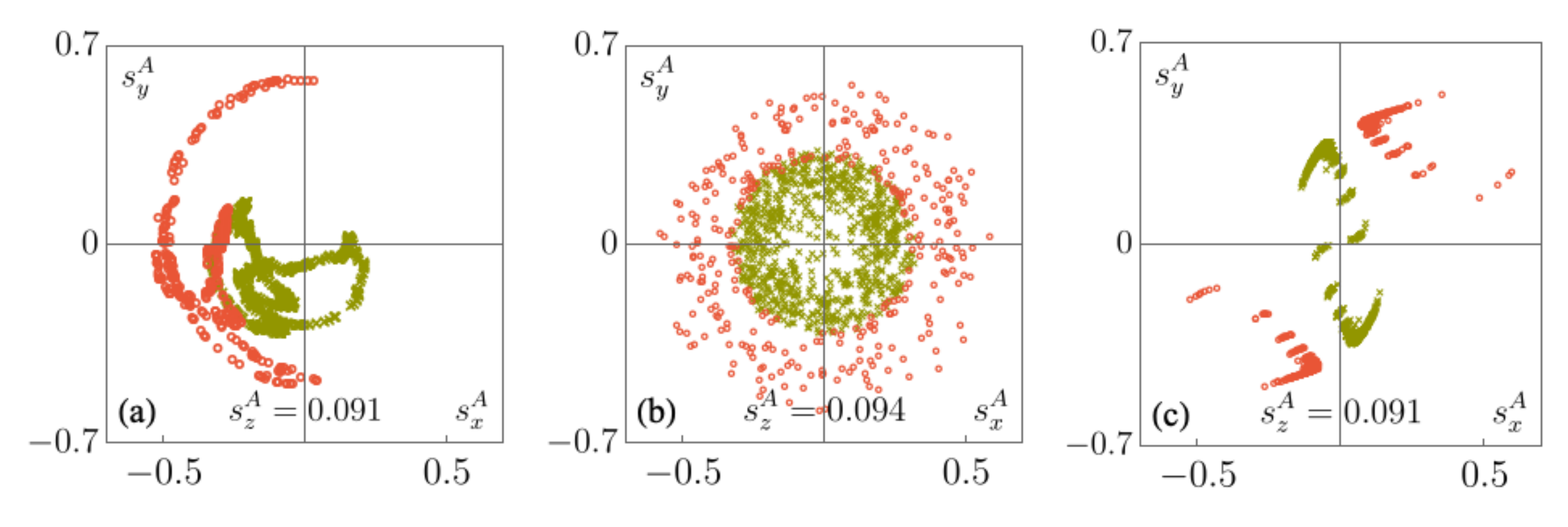}
\begin{picture}(0,0)
\put(-395,90){\includegraphics[height=2.5cm]{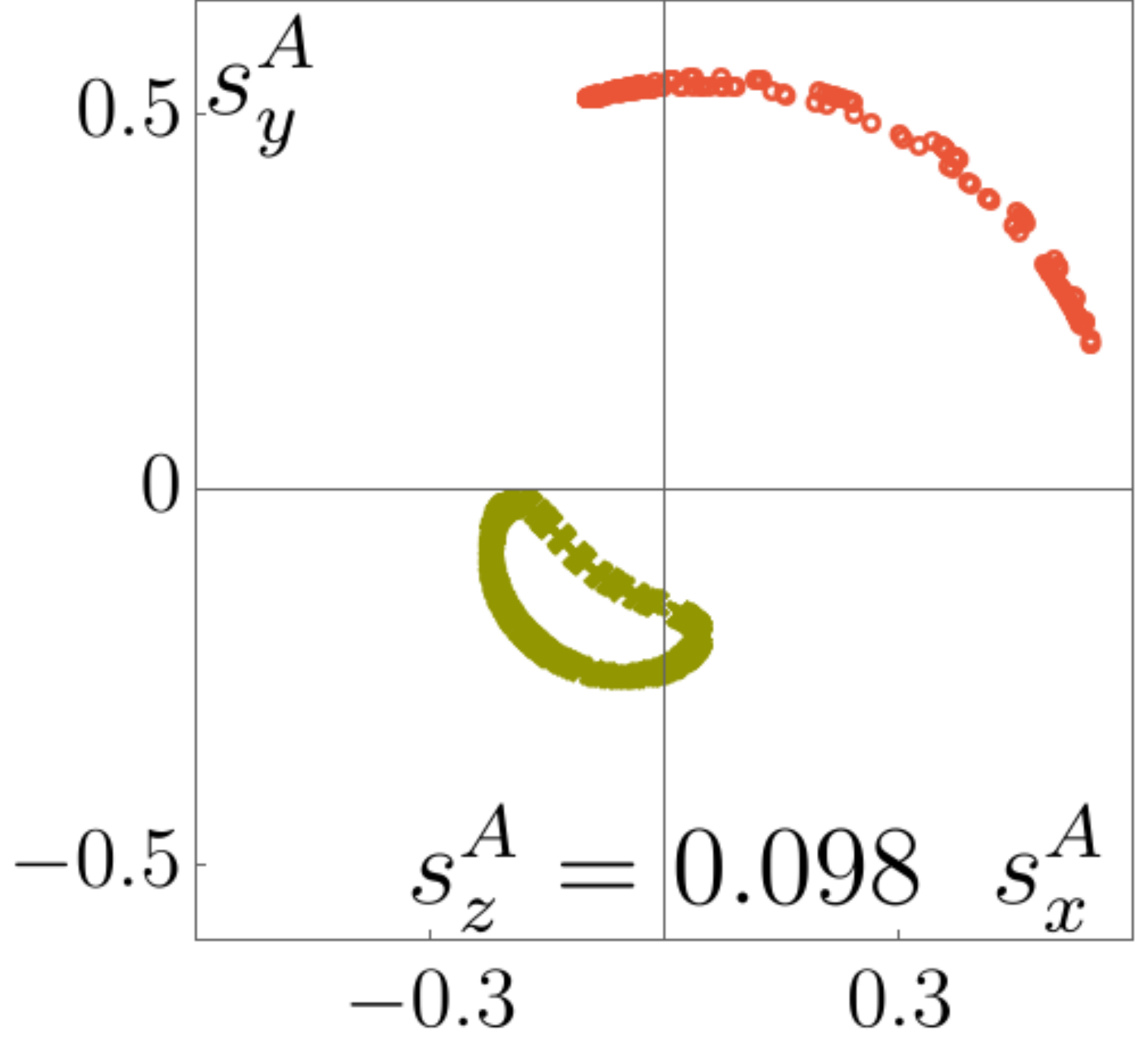}}
\end{picture}
\caption{(color online) Poincar\'{e} sections of the spin $\bm{s}^{A}$  trajectory for (left to right) quasiperiodic ($\delta = 0.115, W = 0.055$), chaotic  ($\delta = 0.1, W = 0.055$) and synchronized chaotic ($\delta = 0.080, W = 0.055$) attractors. These values of $\delta$ and $W$ are the same as in Figs.~\ref{Spectrum_C_Full} and \ref{Cond_Lya}. We section the trajectories with a plane $s^A_z=\mbox{const}$ as explained in the main text.  Orbits cross the plane either from  below (red circles) or from  above (green crosses) generating two distinct Poincar\'{e} sections.    The inset in
\textbf{(a)} shows  an example ($\delta = 0.24, W = 0.055$) of  non-self-intersecting Poincar\'{e} sections of a quasiperiodic attractor. The difference between  Poincar\'{e} sections of chaotic and synchronized chaotic trajectories of spin $\bm{s}^{A}$ is due to the restriction of the dynamics to the synchronization manifold in the latter case.}\label{Poincare}
\end{center}
\end{figure*}

\begin{figure*}[tbp!]
\centering
\subfloat[\;\;\;\;\;\;\;\;\;\;\;\large (a)]{\label{Comp_Traj}\includegraphics[scale = 0.7]{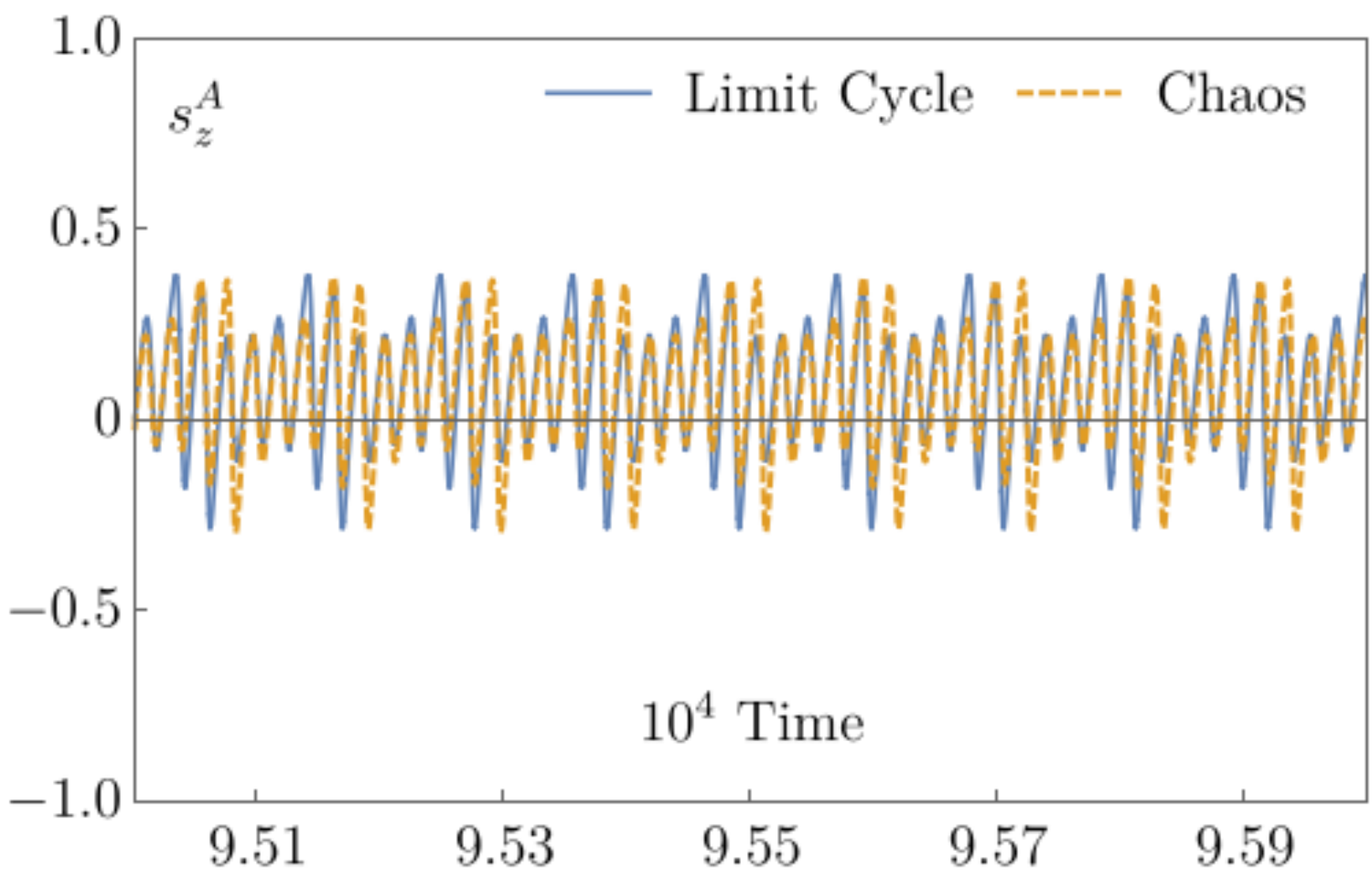}}\\
\subfloat[\;\;\;\;\;\;\;\;\;\;\;\large (b)]{\label{Comp_Lya_Exp}\includegraphics[scale = 0.48]{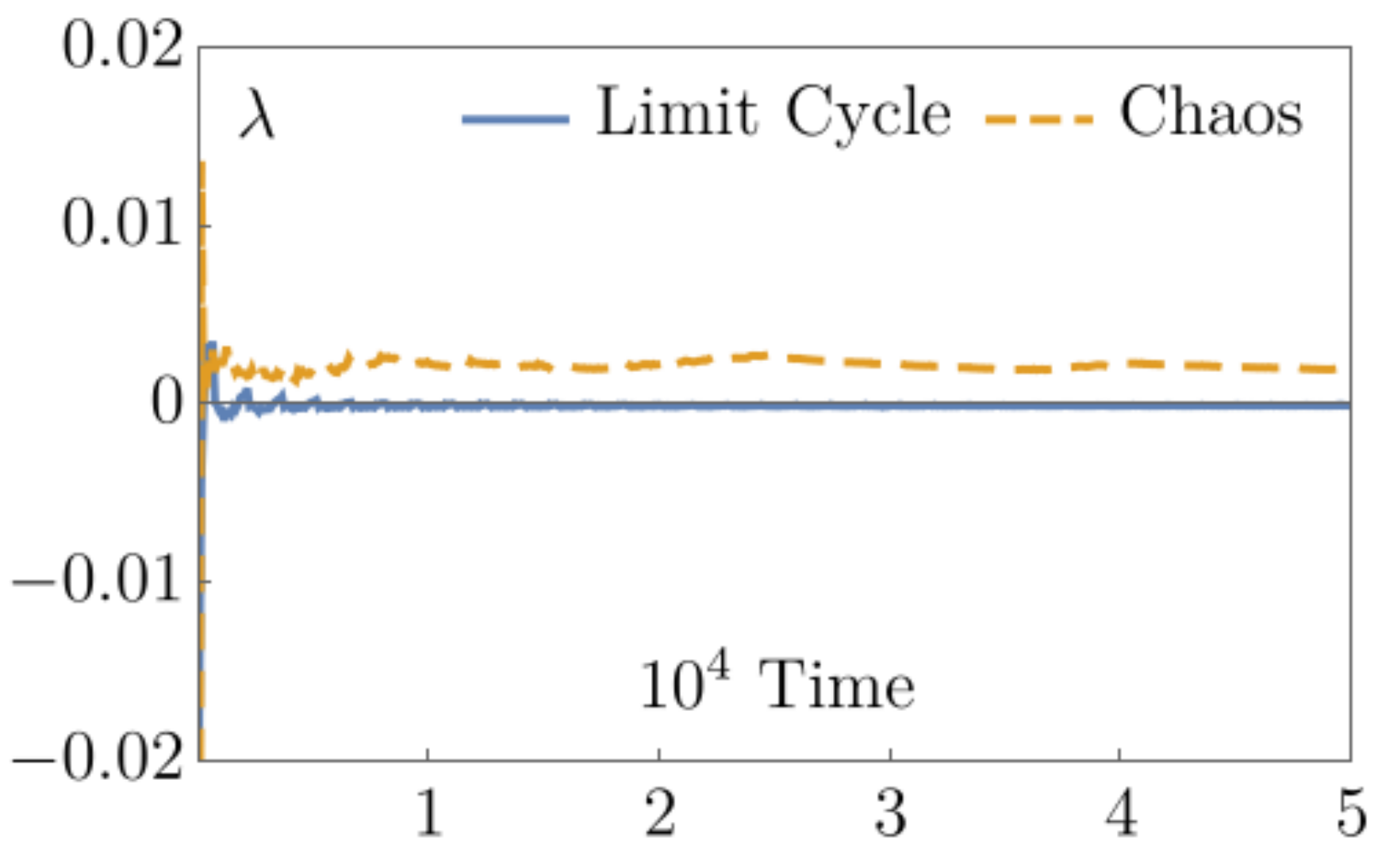}}\qquad
\subfloat[\;\;\;\;\;\;\;\;\;\;\;\large (c)]{\label{Comp_Spec}\includegraphics[scale = 0.46]{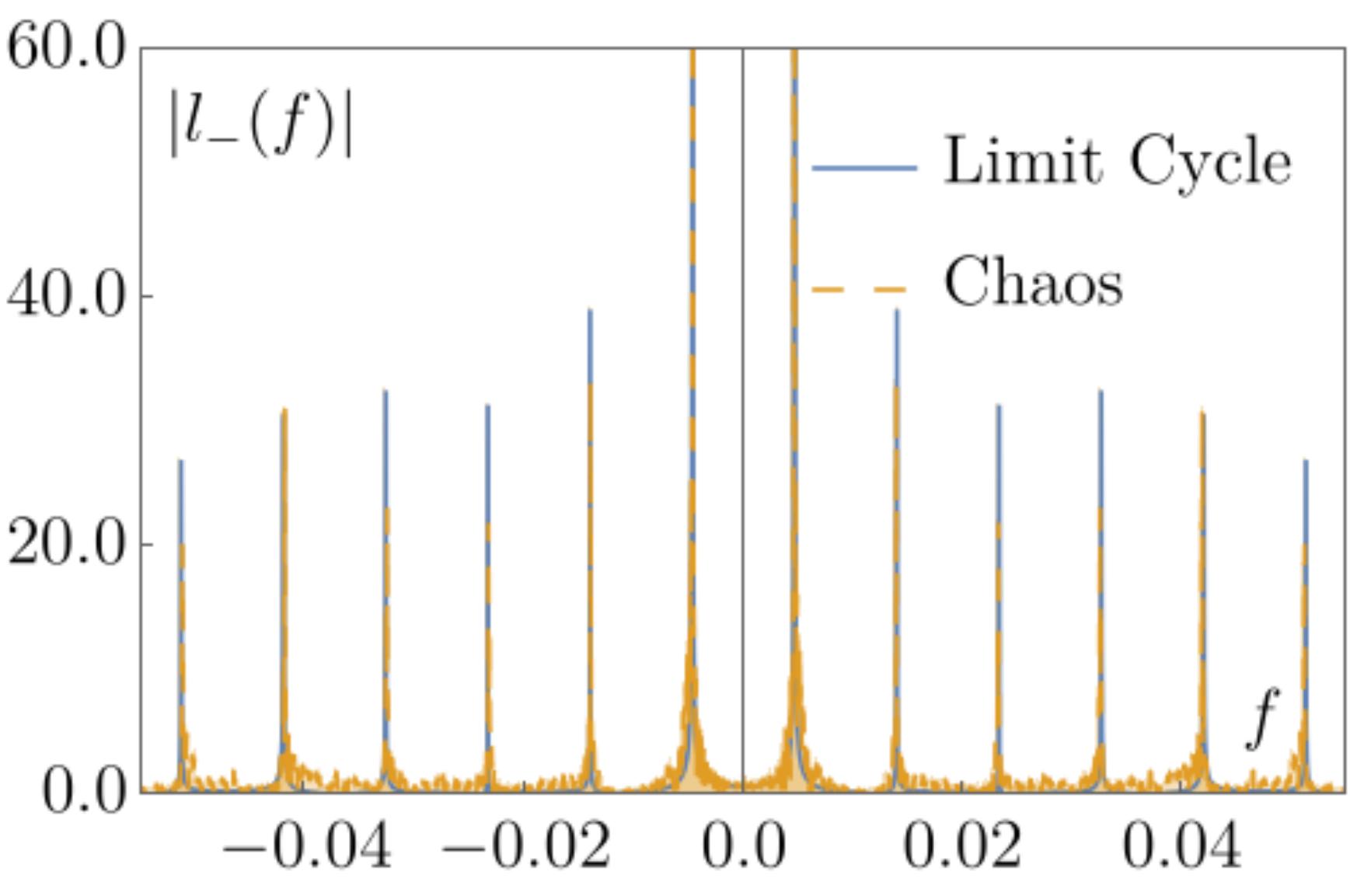}}
\caption{Comparison of $\Z2$-symmetric limit cycle ($\delta = 0.107, W = 0.055$) with synchronized chaos ($\delta = 0.106, W = 0.055$). For $W = 0.055$, the $\Z2$-symmetric limit cycle loses stability in the $\Z2$-symmetric submanifold via tangent bifurcation intermittency between $\delta = 0.107$ and $0.106$. However, note that non-$\Z2$-symmetric  initial conditions  produce quasiperiodicy for both sets of ($\delta, W$). \textbf{(a)}   $s^{A}_{z}(t)$ for the same initial condition. The synchronized chaotic trajectory   closely follows the periodic trajectory. \textbf{(b)} Lyapunov exponents $\lambda$ reveal the chaotic nature ($\lambda > 0$) of the dashed yellow trajectory. \textbf{(c)}  The power spectra.   For the $\Z2$-symmetric limit cycle   the peaks are at $\pm f_{0}, \pm 3f_{0}, \pm 5f_{0, \cdots}$, where $f_{0} \approx 0.0047$. In the chaotic spectrum, the prominent peaks are at the same positions as  for the limit cycle. Nevertheless,   all  frequencies  acquire nonzero weights at the advent of chaos in the $\Z2$-symmetric submanifold.} \label{Comparison_LC_C_Z2}
\end{figure*}

\begin{figure}[tbp!]
\begin{center}
\includegraphics[scale=0.45]{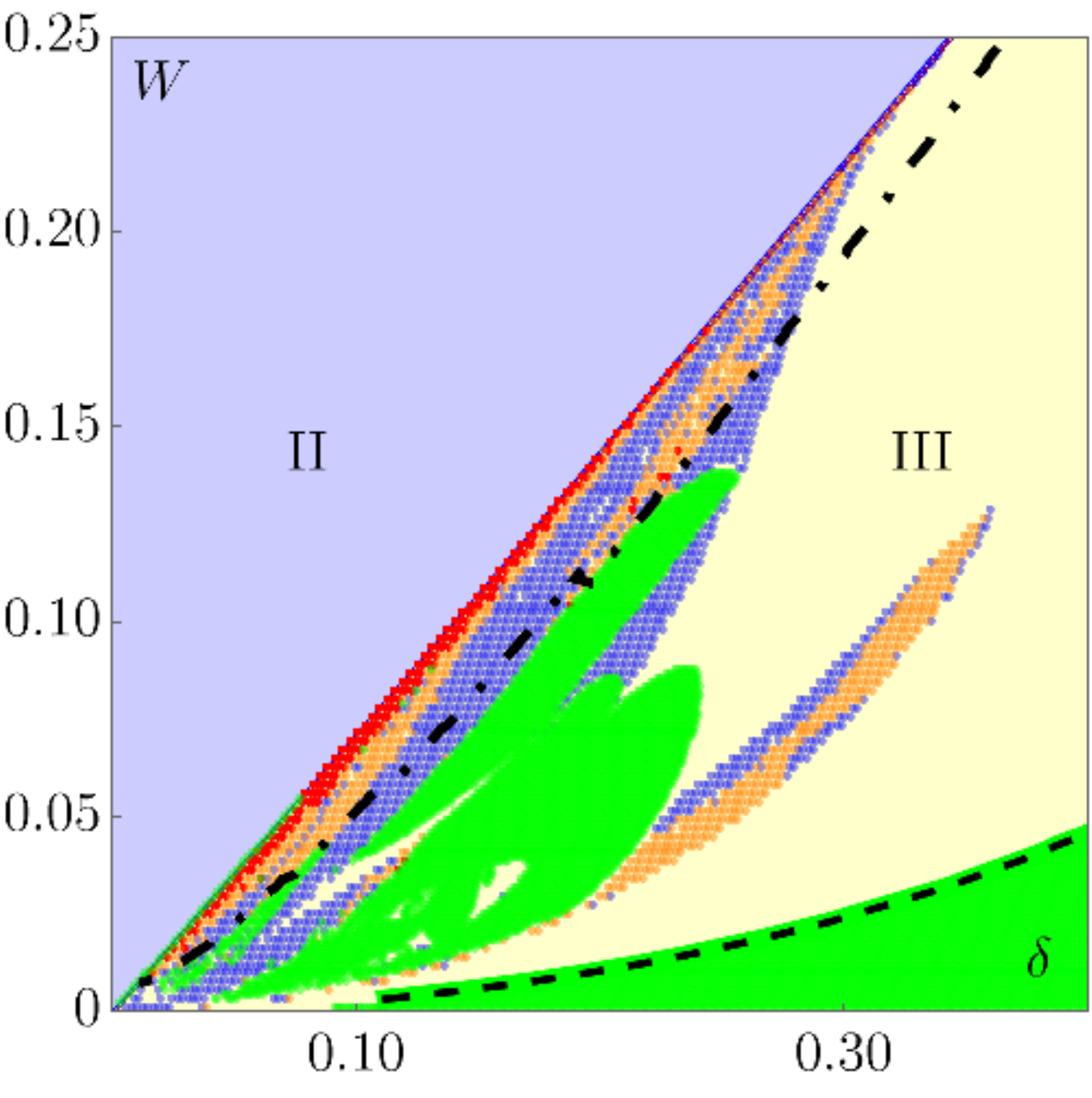}
\caption{(color online) Onset of chaos via tangent bifurcation intermittency in the dynamics confined to the $\Z2$-symmetric submanifold. Initial conditions  lying in this submanifold lead to (synchronized) chaos to the left of the dot-dashed line. We  superimposed this line onto \fref{Near_Origin}. Notice that at its birth the synchronized chaotic attractor is unstable, since only the red points in the immediate vicinity of the II-III boundary produce synchronized chaos for  generic initial conditions  in the full phase-space.}\label{Start_SC}
\end{center}
\end{figure} 

\begin{figure}[tbp!]
\begin{center}
\includegraphics[scale=0.40]{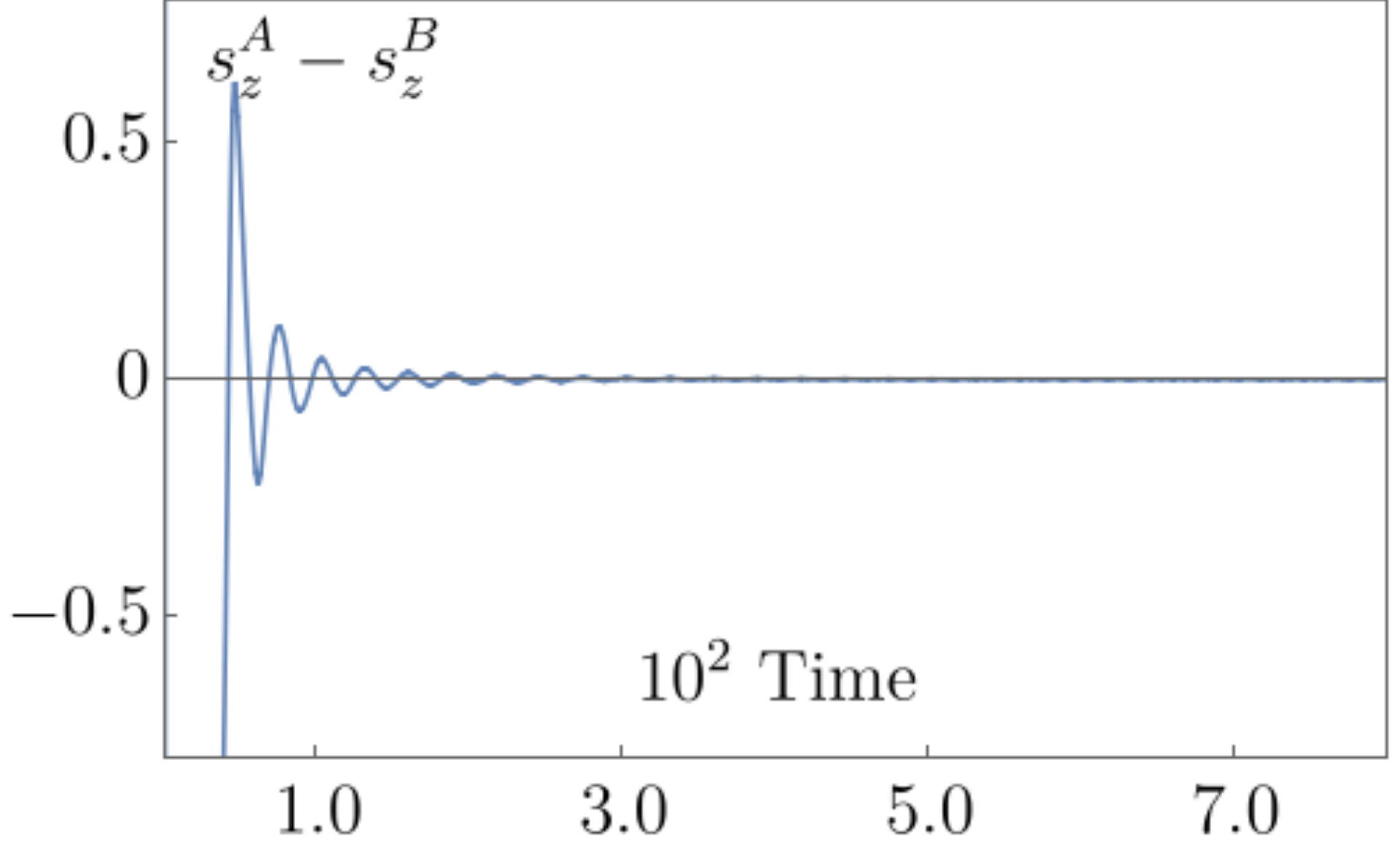}
\caption{The onset of synchronized chaos. The plot
of $s^{A}_{z} - s^{B}_{z}$ vs. time, for $\delta = 0.080$ and $W = 0.055$ shows that the synchronization of chaos is
attained in approximately $200 (N\Gamma_c)^{-1} \approx 0.14$ s.}\label{Onset_SC}
\end{center}
\end{figure} 

\begin{figure}[tbp!]
\begin{center}
\includegraphics[scale=0.40]{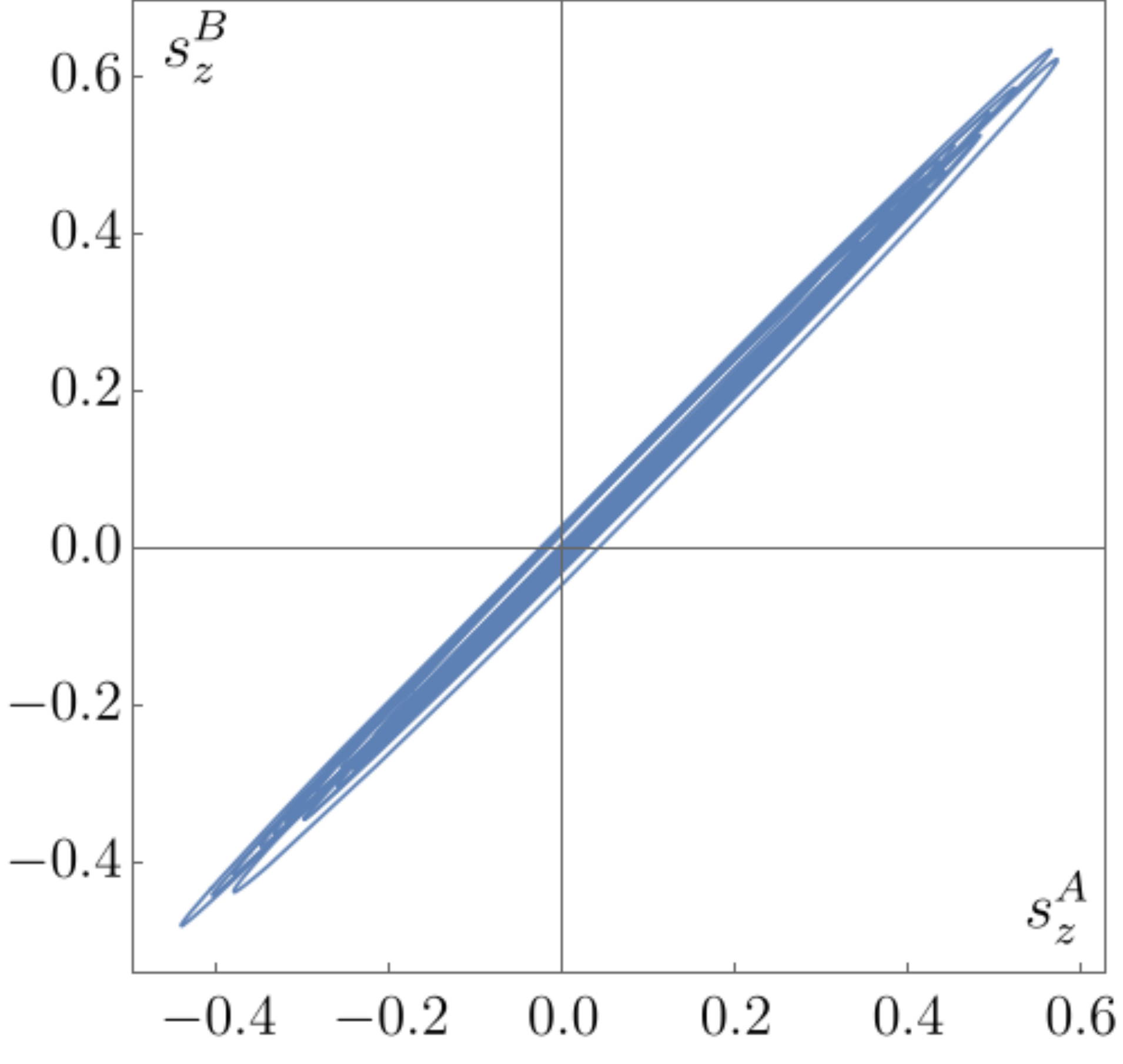}
\caption{The perturbative effect on synchronized chaos of imbalanced atom number between the two ensembles. We plot asymptotic $s_{z}^{B}$ vs. $s_{z}^{A}$ for same $\delta$ and $W$ as in \fref{SC_SZA_vs_SZB}, but $N^{A}/N_{\textrm{av}} = 0.95$ as opposed to $N^{A}/N_{\textrm{av}} = 1$ in \fref{SC_SZA_vs_SZB}. Here $N_{\textrm{av}} = (N^{A} + N^{A})/2$.}\label{Atom_No_Imbalance}
\end{center}
\end{figure} 

A Poincar\'{e} section of an attractor is the set of points where its trajectory crosses a   plane cutting the attractor into two,  counting only the crossings that occur in one direction~\cite{Poincare_Paper, Hilborn}. To obtain a 2D representation,  we show  the Poincar\'{e} sections for the A spin only  in \fref{Poincare}.  Those for the B spin are qualitatively similar.  We cut the trajectory of the A spin with the plane    $s^{A}_{z} = \textrm{const} = \frac{1}{t_1}\int_{t_{0}}^{t_{0} + t_1}\!\! s^{A}_{z} dt$ parallel to the $s_{x}^{A} - s_{y}^{A}$ plane, where $t_{0}$ and $t_1$ are sufficiently large.  

Poincar\'{e} sections  of quasiperiodic trajectories appear as continuous curves.  Consider, e.g., a two-frequency quasiperiodic motion. It occurs on a 2D torus in  
the 6D  space of six components of both classical spins.
 We expect the full Poincar\'e section to be  a closed non-self-intersecting 5D curve. However, when looking   only at the A spin, we project this curve onto a 2D plane.  The resulting Poincar\'{e} section is still a continuous curve, but it can now  intersect itself as  in the leftmost plot  in \fref{Poincare}.   The Poincar\'{e} section  of a chaotic trajectory appears as a smudge of random points. Finally,  the  section of a chaotic synchronized trajectory is a collection of disjoint segments highlighting both the chaotic and   constricted (to the synchronization manifold) nature of the dynamics. 

An experimentally observable quantity is the power spectrum, $|\bm{E}(f)|^2$, of the light emitted by the cavity \citep{Patra_1, Carmichael_3}. Here $\bm{E}(f)$   is the Fourier transform  of the  (complex) radiated electric field and $f$ is the frequency. Within the mean-field approximation, we find  $|\bm{E}(f)|^2\propto |l_{-}(f)|^{2}$, where  $l_-=l_x-\imath l_y$ and $\bm{l}$ is the total classical spin.
 The quasiperiodic spectrum (\fref{Spectrum_QP}) has main peaks at $0, \pm f_{1}, \pm 2f_{1}, \cdots$, with auxiliary peaks  spaced at     $f_{2}$ bunched around them. We did not observe more than two-frequency quasiperiodicity. While it is generally difficult to differentiate between chaotic and    quasiperiodic spectra \cite{QP_Chaos_Spectra},  in our system     the latter are visibly  discrete.  Nevertheless, we do not rely on this feature and use maximum Lyapunov exponents   to  distinguish quasiperiodicity and chaos in \fref{Near_Origin}. Note that although the chaotic spectrum is continuous, it features distinct peaks that are independent of the initial conditions (\fref{Spectrum_C}).  In contrast to the spectrum of the synchronized chaotic attractor in \fref{Spectrum_SC}, both chaotic and quasiperiodic power spectra have prominent peaks at the origin and  no reflection symmetry.

Near the boundary between Phases II and III in \fref{Near_Origin}  (red points), we observe synchronized chaos. Since the dynamics in this subregion restore the $\Z2$ symmetry, the solutions are confined to the 4D $\Z2$-symmetric submanifold. We write the two constraint relations (independent of initial condition) as  
\beg 
\begin{split} 
\big(s_{x}^{A}\big)^{2} + \big(s_{y}^{A}\big)^{2} = \big(s_{x}^{B}\big)^{2} + \big(s_{y}^{B}\big)^{2}\!\!,\quad s_{z}^{A} = s_{z}^{B}.
\end{split} 
\label{Chaotic_Synch_Synch_Sub-manifold}%
\en%
For our purposes, these relations define the synchronization manifold.  Coordinates spanning the ``transverse manifold" (complementary to the synchronization manifold) are
\beg 
\begin{split} 
\mathsf n_{1} \equiv  \big(s_{x}^{A}\big)^{2} + \big(s_{y}^{A}\big)^{2} -  \big(s_{x}^{B}\big)^{2} - \big(s_{y}^{B}\big)^{2}\!\!,\\
\mathsf n_{2} \equiv s_{z}^{A} - s_{z}^{B}.
\end{split} 
\label{Trans_Coord}%
\en%
We derive the evolution equations  for the transverse subsystem from \eref{Mean-Field_1} as
\begs
\bea
\dot{\mathsf{n}}_{1} &=& \frac{1}{2}\big(l_{z} - 2W\big)\mathsf n_{1} + \frac{1}{2}\big(l_{x}^{2} + l_{y}^{2}\big)\mathsf n_{2}, \\[0pt] 
\dot{\mathsf{n}}_{2} &=& -\frac{\mathsf  n_{1}}{2}-W\mathsf n_{2}.
\eea 
\label{Transverse_Coordinates_Eqn}%
\ens%
To compute the  conditional Lyapunov exponent for an attractor, we first determine its $l_z$ and $l_{x}^{2} + l_{y}^{2}$ with the help of \eref{Mean-Field_1}. These serve as time-dependent coefficients in \esref{Transverse_Coordinates_Eqn}.  In principle, we should linearize \esref{Transverse_Coordinates_Eqn} in small deviations $\Delta\mathsf n_{1}$ and $\Delta\mathsf n_{2}$. However, since these equations are already linear, we simply redefine $\mathsf n_{1}$ and $\mathsf n_{2}$ to be such arbitrary infenitesimal deviations in transverse directions and numerically simulate \esref{Transverse_Coordinates_Eqn}.
The conditional Lyapunov exponent is the rate of growth of distances in the transverse manifold, i.e., it is given by \eref{lyapunov}, where $d=\sqrt{\mathsf n_1^2+\mathsf n_2^2}$ is the transverse distance.
For chaotic synchronized trajectories $\lambda_{c} \approx -10^{-2} \pm 10^{-5}$ for $t\ge 5 \times 10^{4}$, whereas for chaotic ones $\lambda_{c}\approx \pm 10^{-5}$ (\fref{Cond_Lya}). On the other hand,  maximum Lyapunov exponents $\lambda$ for both chaos and synchronized chaos behave similarly. 

In \aref{sec:Tan_Bifurc_Intmit},  we explain the emergence of the synchronized chaos in our system   via \textit{tangent bifurcation intermittency} of the $\Z2$-symmetric limit cycle. As a result, synchronized chaotic trajectories spend most of their time in the vicinity of the now unstable $\Z2$-symmetric limit cycle, see \fref{Comp_Traj}.  Further, we observe in \fref{Start_SC} that the synchronized chaotic attractor starts off being unstable in the full phase space. Only close to the boundary of the Phases II and III (red points) this attractor becomes sufficiently attractive \cite{Chaotic_Synch_X}.   The restoration of the $\Z2$ symmetry explains the reflection symmetric (with no peak at zero) power spectrum of the synchronized  chaotic attractor, see \fsref{Spectrum_SC}.

In the master equation \re{Full_Master} we  neglected the effects of spontaneous emission and inhomogeneous life time $T_{2}$. Since the chaotic synchronization is an asymptotic solution of  mean-field equations \re{Mean-Field_1}, one needs to clarify the effects of these neglected decay processes at large times. To this effect, we show in \fref{Onset_SC} that the system reaches its steady state for $\delta = 0.080$ and $W = 0.055$ (same as in \fsref{SC_ZA}, \ref{SC_SZA_vs_SZB}, and \ref{Spectrum_SC}) in approximately $0.14$~s ($\approx 200$ time steps). In a typical experimental setup the timescale related to spontaneous emission can be pushed to $100$~s, whereas $T_{2}$ can be as large as $1$~s~\cite{Holland_One_Ensemble_Theory_1}. This comparison of the experimental timescales with the timescale relevant for the observation of chaotic synchronization validates the master equation \re{Full_Master}.

Another impediment for the observation of chaotic synchronization is the limited efficiency of the atomic traps that are required to localize the atomic ensembles. Current experiments \cite{Holland_One_Ensemble_Expt_1} are able to observe superradiant emission for as long as $120$~ms. This should be enough to detect the signature of chaotic synchronization -- exponential attenuation of $s^{A}_{z} - s^{B}_{z}$. For example, in \fref{Onset_SC} this is seen between $70$ and $140$~ms. Loss of atoms from the traps with different rates can also lead to atom number imbalance between the ensembles, which in turn breaks the $\Z2$ symmetry. This effect, however, affects the steady states only perturbatively, see \fref{Atom_No_Imbalance}.

\section{Conclusion}

In conclusion, we have  predicted chaotic synchronization of the dynamics of two atomic ensembles collectively coupled to a heavily damped cavity mode. Synchronized chaos  emerges from quasiperiodicity by way of (asymmetric) chaos. Its origin is in the tangent bifurcation intermittency of the $\Z2$-symmetric limit cycle (see \aref{sec:Tan_Bifurc_Intmit}).   We distinguish the three phases theoretically, by analyzing the Poincar\'{e} sections and maximum   and   conditional Lyapunov exponents. Open questions include the effects of coupling to multiple cavity modes and of quantum fluctuations. A quantum analogue to our system, where the overall system is chaotic, but a subsector is not, is known \cite{Qtm_Chaos}.  It would also be  interesting to explore prospects of realizing a viable steganography \cite{Uchida, Chaotic_Synch_1, Chaotic_Synch_2, Chaotic_Synch_3, Chaotic_Synch_6} (instead of hiding the meaning of transmitted message, hide the existence of the message itself) protocol with our system. In particular, it  is not apparent how to send a message over a long distance.

This work was   supported by the National Science Foundation Grant DMR1609829.

\appendix

\section{Tangent Bifurcation Intermittency in the $\Z2$-symmetric Submanifold}
\label{sec:Tan_Bifurc_Intmit}

 Recall that the synchronized chaotic attractor spontaneously restores the $\Z2$ symmetry between the two ensembles of atoms. 
 Therefore, to gain further insight into  it, we investigate    $\Z2$-symmetric dynamics in this section. 
 
 After a rotation
 around the z-axis by an angle $\phi_0$, which depends on the initial condition,  $\Z2$-symmetric dynamics are invariant with respect to the replacement \re{Z2}, i.e., 
 \beg
 s_x^B=s_x^A,\quad s_y^B=-s_y^A, \quad s_z^B=s_z^A.
 \label{z2sup}
 \en
This implies $l_{x} = 2s^{A}_{x} = 2s^{B}_{x}$, $l_{y} = 0$ and the mean-field equations of motion \re{Mean-Field_1} for the spin $\bm{s}^{A}$ become
\begs
\bea
\dot{s}_{x} &=& - \frac{\delta}{2}s_{y}- \frac{W}{2}s_{x} + s_{z}s_{x},\label{ReducedX} \\ 
\dot{s}_{y} &=& \frac{\delta}{2}s_{x} - \frac{W}{2}s_{y}, \label{ReducedY}\\ 
\dot{s}_{z} &=&  W\big(1 - s_{z}\big) -  s_{x}^{2}, \label{ReducedZ} 
\eea 
\label{Symm_One_Spin_Eqn}%
\ens%
where we dropped the superscript A for simplicity. The spin $\bm{s}^{B}$ is related to $\bm{s}^{A}$ by \eref{z2sup}.  

Note
that \eref{Symm_One_Spin_Eqn} is very different from the mean-field equations of motion for a single atomic ensemble coupled to a bad cavity. We obtain the latter from the two-ensemble   equations~\re{Mean-Field_1} by setting one of the spins, say
$\bm s^B$, to zero. Then, by going into a frame uniformly rotating  with  frequency $\omega_A=\delta/2$, we eliminate $\delta$ from the one-ensemble equations of motion. Thus, single ensemble (one spin) equations correspond to setting $\delta=0$ in \eref{Mean-Field_1}. Indeed, summing \eref{Mean-Field_1} for $\delta=0$ over $\tau$ and rescaling $\bm l\to 2 \bm l$, $W\to 2W$ and $2t\to t$ we obtain the one ensemble equations
of motion in the rotating frame.
This implies that the   nonequilibrium phase diagram for a single ensemble  is just the vertical, $\delta=0$ axis of the two-ensemble phase diagram in \fref{Near_Origin} with the rescaling $2W\to W$. It  consists of two fixed points (normal phase and monochromatic superradiance), see \Rref{Patra_1} for details. In contrast, \eref{Symm_One_Spin_Eqn} depends on two dimensionless parameters $\delta$ and $W$ in an essential way and, consequently, has much richer dynamics as we now discuss.

 While solutions of \eref{Symm_One_Spin_Eqn} are consistent with the full mean-field equations  \re{Mean-Field_1}, their stability  in the full phase space  is not guaranteed. The three types of solutions of \eref{Symm_One_Spin_Eqn} are: fixed points, periodic ($\Z2$ symmetric limit cycle), and chaotic (synchronized chaos). In the parentheses we mention the equivalent solutions of \eref{Mean-Field_1}. 

 Consider the portion of the phase diagram  to the left of   the $\Z2$ symmetry breaking line (black dashed line) in \fref{Start_SC}. Although the $\Z2$-symmetric limit cycle loses stability  in the full phase-space, it is still stable on the $\Z2$-symmetric submanifold.  As we move towards the Phase II-III  boundary, the periodic solution eventually loses  stability on the dot-dashed line in \fref{Start_SC} even on the $\Z2$-symmetric submanifold, giving rise to chaos.  We determine this line in   by computing the maximum Lyapunov exponent $\lambda$ with the help of \esref{Floquet_Eqn_Z2} and~(\ref{Symm_One_Spin_Eqn}). To the left of the dot-dashed line $\lambda>0,$ see, e.g.,  \fref{Comp_Lya_Exp}. Additionally, we prove the loss of stability employing Floquet analysis of \eref{Symm_One_Spin_Eqn}.

The abrupt transition, and the proximity of the periodic and  chaotic attractors suggest \textit{tangent bifurcation intermittency}   \cite{Hilborn}. To illustrate this closeness we compare a chaotic spectrum with an adjacent periodic one in \fref{Comp_Spec}. Below, we provide the final proof in support of this claim by studying the evolution of the Floquet multipliers.

\subsection{Floquet Analysis}\label{subsec:Floquet}

Our goal is to analyze the stability of the periodic solutions of the reduced  equations \re{Symm_One_Spin_Eqn}. To that end, we first summarize the Floquet analysis. Write the solution of \eref{Symm_One_Spin_Eqn} as $\bm{s} + \Delta \bm{s}$, where $\bm{s}$ is the periodic solution with period $T$, and $\Delta \bm{s}$ is a  perturbation. Linearizing \eref{Symm_One_Spin_Eqn} with respect to the perturbation, we obtain a set of linear equations with time-dependent coefficients
\begs
\bea
\frac{\mathrm{d}\Delta s_{x}}{\mathrm{d}t} &=& \big(s_{z} - \frac{W}{2}\big)\Delta s_{x} - \frac{\delta}{2}\Delta s_{y} + s_{x}\Delta s_{z}, \\ [7pt]
\nonumber\\%
\frac{\mathrm{d}\Delta s_{y}}{\mathrm{d}t} &=& \frac{\delta}{2}\Delta s_{x} - \frac{W}{2}\Delta s_{y}, \\[7pt] 
\frac{\mathrm{d}\Delta s_{y}}{\mathrm{d}t} &=& -2s_{x}\Delta s_{x} - W\Delta s_{z}. 
\eea 
\label{Floquet_Eqn_Z2}
\ens%
 The next step is to determine the monodromy matrix $\mathbb{M}=[\mathbb{S}(0)]^{-1} \mathbb{S}(T)$ for \eref{Floquet_Eqn_Z2}. Here
 $\mathbb{S}(t)$ is a $3\times3$ matrix. Its  columns are any three linearly independent solutions of \eref{Floquet_Eqn_Z2}, which we obtain numerically. The eigenvalues of the monodramy matrix $\rho_i\equiv e^{\varkappa_i T}$ are known as Floquet multipliers  and $\varkappa_i$ are the corresponding Floquet exponents.  
 By Floquet theorem, the general solution of \eref{Floquet_Eqn_Z2} is  
 \beg
 \Delta \bm{s}(t) =\sum_{i=1}^3 C_i e^{\varkappa_i t} \bm{p}_i(t),\quad \rho_i\equiv e^{\varkappa_i T},
 \label{Flthm}
 \en
where $C_i$ are constants and $\bm{p}_i(t)$ are linearly independent and periodic with period $T$ vectors. The limit cycle looses stability when the absolute value of one of the Floquet multipliers becomes greater than one.

Further, notice that $\Delta \bm{s} = \dot{\bm{s}}$ is a purely periodic with period $T$ solution of \eref{Floquet_Eqn_Z2}. This implies that one
of the Floquet multipliers is identically equal to one, so that \eref{Flthm} takes the form
\beg 
\Delta \bm{s}(t) = C_{1}\dot{\bm s}(t)+ C_{2}e^{\varkappa_{2}t}\bm{p}_{2}(t) + C_{3}e^{\varkappa_{3}t}\bm{p}_{3}(t). 
\label{Floquet_Gen_Perturb}
\en%
 Near the dot-dashed line in \fref{Start_SC},   the remaining Floquet multipliers $\rho_{2}$ and $\rho_3$ are both real. As we approach this line from the right, $|\rho_{2}|$ tends to one from below, while $|\rho_{3}|$ remains less than one across criticality.  
 
 This behavior of the Floquet multipliers  implies (by definition) tangent bifurcation intermittency route to chaos \cite{Hilborn}. A key feature of this route to chaos is that the chaotic attractor  right after the bifurcation remains close to the now unstable limit cycle for most of the time, which we indeed observe in  Figs.~\ref{Comp_Traj} and    \ref{Comp_Spec}. Further, the power spectrum of the $\Z2$-symmetric limit cycle is known to have reflection symmetry and no peak at zero frequency~\cite{Patra_1}.   The power spectrum of the synchronized chaotic attractor in  \fref{Comp_Spec}  reproduces 
 these features  due to the proximity of its trajectory to the limit cycle.

%


\begin{thebibliography}{999}

\bibitem{Uchida} A. Uchida,  \textit{Optical Communication with Chaotic Lasers} (Wiley-VCH, 2012).

\bibitem{Chaotic_Synch_1} L. M. Pecora, T. L. Carroll, G. A. Johnson, and D. G. Mar, Fundamentals of Synchronization in Chaotic Systems, Concepts and Applications, \href{https://aip.scitation.org/doi/10.1063/1.166278}{Chaos \textbf{7}, 520 (1997)}.

\bibitem{Chaotic_Synch_2} L. M. Pecora and T. L. Carroll, Synchronization in Chaotic Systems, \href{https://journals.aps.org/prl/abstract/10.1103/PhysRevLett.64.821}{Phys. Rev. Lett. \textbf{64}, 821 (1990)}.

\bibitem{Chaotic_Synch_3} S. Hayes, C. Grebogi, E. Ott and A. Mark, Experimental Control of Chaos for Communication, \href{https://journals.aps.org/prl/abstract/10.1103/PhysRevLett.73.1781}{Phys. Rev. Lett. \textbf{73}, 1781 (1994)}.

\bibitem{Chaotic_Synch_6} P. M. Alsing, A. Gavrielides, V. Kovanis, R. Roy and K. S. Thornburg, Jr., Encoding and Decoding Messages with Chaotic Lasers, \href{https://journals.aps.org/pre/abstract/10.1103/PhysRevE.56.6302}{Phys. Rev. E. \textbf{56}, 6302 (1997)}.

\bibitem{Chaotic_Synch_4} H. G. Winful and L. Rahman, Synchronized Chaos and Spatiotemporal Chaos in Array of Coupled Lasers, \href{https://journals.aps.org/prl/abstract/10.1103/PhysRevLett.65.1575}{Phys. Rev. Lett. \textbf{65}, 1575 (1990)}.

\bibitem{Chaotic_Synch_5} R. Roy and K. S. Thornburg, Jr., Experimental Synchronization of Chaotic Lasers, \href{https://journals.aps.org/prl/abstract/10.1103/PhysRevLett.72.2009}{Phys. Rev. Lett. \textbf{72}, 2009 (1994)}.



\bibitem{Chaotic_Synch_7} T. Fukuyama, R. Kozakov, H. Testrich and C. Wilke, Spatiotemporal Synchronization of Coupled Oscillators in a Laboratory Plasma, \href{https://journals.aps.org/prl/abstract/10.1103/PhysRevLett.96.024101}{Phys. Rev. Lett. \textbf{96}, 024101 (2006)}.

\bibitem{Chaotic_Synch_8} B. Blasius, A. Huppert, and L. Stone, Complex dynamics and phase synchronization in spatially extended ecological systems, \href{https://www.nature.com/articles/20676}{Nature \textbf{399}, 354 (1999)}.

\bibitem{Chaotic_Synch_9} M. de Sousa Vieira, Chaos and Synchronized Chaos in an Earthquake Model, \href{https://journals.aps.org/prl/abstract/10.1103/PhysRevLett.82.201}{Phys. Rev. Lett. \textbf{82}, 201 (1999)}.

\bibitem{Holland_Two_Ensemble_Expt} J. M. Weiner, K. C. Cox, J. G. Bohnet, and J. K. Thompson, Phase synchronization inside a superradiant laser, \href{https://journals.aps.org/pra/abstract/10.1103/PhysRevA.95.033808}{Phys. Rev. A \textbf{95}, 033808 (2017)}.

\bibitem{Holland_atomic_clock} Minghui Xu and M. J. Holland, Conditional Ramsey Spectroscopy with Synchronized Atoms, \href{https://journals.aps.org/prl/abstract/10.1103/PhysRevLett.114.103601}{Phys. Rev. Lett. \textbf{114}, 103601 (2015)}.

\bibitem{Holland_Two_Ensemble_Theory} Minghui Xu, D. A. Tieri, E. C. Fine, James K. Thompson and M. J. Holland, Synchronization of Two Ensembles of Atoms, \href{https://journals.aps.org/prl/abstract/10.1103/PhysRevLett.113.154101}{Phys. Rev. Lett. \textbf{113}, 154101 (2014)}.

\bibitem{Patra_1} A. Patra, B. L. Altshuler and E. A. Yuzbashyan,  Driven-Dissipative Dynamics of  Atomic Ensembles in a Resonant Cavity: Nonequilibrium Phase Diagram and Periodically Modulated Superradiance, \href{https://journals.aps.org/pra/abstract/10.1103/PhysRevA.99.033802}{Phys. Rev. A \textbf{99}, 033802 (2019)}.

\bibitem{Carmichael_3} H. J. Carmichael, \textit{An Open System Approach to Quantum Optics} (Springer-Verlag Berlin Heidelberg, 1993).

\bibitem{Poincare_Paper} W Tucker, Computing Accurate Poincar\'{e} Maps, \href{https://www.sciencedirect.com/science/article/pii/S0167278902006036}{Physica D \textbf{171}, 127 (2002)}.

\bibitem{Hilborn} R. C. Hilborn, \textit{Chaos and Nonlinear Dynamics, An Introduction for Scientists and Engineers, Second Edition} (Oxford University Press, 2001).

\bibitem{QP_Chaos_1} D. Ruelle and F. Takens, On the Nature of Turbulence, \href{https://link.springer.com/article/10.1007/BF01646553}{Commun. Math. Phys. \textbf{20}, 167 (1971)}.

\bibitem{QP_Chaos_2} S. E. Newhouse, D. Ruelle and F. Takens, Occurance of Strange Axiom A Attractors near Quasi-periodic Flows on $T^{m}, m\geq 3$, \href{https://link.springer.com/article/10.1007/BF01940759}{Commun. Math. Phys. \textbf{64}, 35 (1978)}.


\bibitem{Bad_cavity} R. Bonifacio, P. Schwendimann, and Fritz Haake, Quantum Statistical Theory of Superradiance. I, \href{https://journals.aps.org/pra/abstract/10.1103/PhysRevA.4.302}{Phys. Rev. A \textbf{4}, 302 (1971)}.


\bibitem{QP_Chaos_Spectra} R. S. Dumont and P. Brumer, Characteristics of power spectra for regular and chaotic systems, \href{https://aip.scitation.org/doi/10.1063/1.454126}{J. Chem. Phys. \textbf{88}, 1481 (1988)}.

\bibitem{Chaotic_Synch_X} K. Josi\'{c}, Invariant Manifolds and Synchronization of Coupled Dynamical Systems, \href{https://journals.aps.org/prl/abstract/10.1103/PhysRevLett.80.3053}{Phys. Rev. Lett. \textbf{80}, 3053 (1998)}.

\bibitem{Holland_One_Ensemble_Theory_1} D. Meiser, Jun Ye, D. R. Carlson and M. J. Holland, Prospects for a Milihertz-
Linewidth Laser, \href{https://journals.aps.org/prl/abstract/10.1103/PhysRevLett.102.163601}{Phys. Rev. Lett. \textbf{102}, 163601 (2006)}.

\bibitem{Holland_One_Ensemble_Expt_1} J. G. Bohnet, Z. Chen, J. M. Weiner, D. Meiser, M. J. Holland
and J. K. Thompson, A steady-state superradiant laser with less than one
intracavity photon, \href{https://www.nature.com/articles/nature10920}{Nature \textbf{484}, 78 (2012)}.

\bibitem{Qtm_Chaos} V. A. Yurovsky, Long-lived states with well-defined spins in 
spin $- 1/2$ homogeneous Bose gases, \href{https://journals.aps.org/pra/abstract/10.1103/PhysRevA.93.023613}{Phys. Rev. A \textbf{93}, 023613 (2016)}.



 
\end{thebibliography}
\end{document}